\documentclass[12pt,english,12pt,aps,prd,manuscript,superscriptaddress]{revtex4-1}
\usepackage[T1]{fontenc}
\usepackage[latin9]{inputenc}
\usepackage{color}
\usepackage{babel}
\usepackage{amstext}
\usepackage{graphicx}
\usepackage{esint}
\usepackage[unicode=true,pdfusetitle,
 bookmarks=true,bookmarksnumbered=false,bookmarksopen=false,
 breaklinks=false,pdfborder={0 0 1},backref=false,colorlinks=true]
 {hyperref}
\hypersetup{
 pdfborderstyle=,linkcolor=blue,citecolor=cyan}

\makeatletter

\providecommand{\tabularnewline}{\\}

\usepackage{babel}

\usepackage{xcolor}

\makeatother

\begin{document}
\title{Space-average electromagnetic fields and EM anomaly weighted by energy
density in heavy-ion collisions}
\author{Irfan Siddique}
\email{irfans@mail.ustc.edu.cn}

\affiliation{Key Laboratory of Particle Physics and Particle Irradiation (MOE),
Institute of Frontier and Interdisciplinary Science, Shandong University,
Qingdao, Shandong 2666237, China.}
\author{Xin-Li Sheng}
\email{xls@mail.ustc.edu.cn}

\affiliation{Key Laboratory of Quark and Lepton Physics (MOE) and Institute of
Particle Physics, Central China Normal University, Wuhan, Hubei 430079,
China.}
\author{Qun Wang}
\email{qunwang@ustc.edu.cn}

\affiliation{Interdisciplinary Center for Theoretical Study and Department of Modern
Physics, University of Science and Technology of China, Hefei 230026,
China.}
\affiliation{Peng Huanwu Center for Fundamental Theory, Hefei, Anhui 230026, China.}
\preprint{USTC-ICTS/PCFT-21-22}
\begin{abstract}
We study the space-average electromagnetic (EM) fields weighted by
the energy density in the central regions of heavy ion collisions.
These average quantities can serve as a barometer for the magnetic-field
induced effects such as the magnetic effect, the chiral separation
effect and the chiral magnetic wave. Comparing with the magnetic fields
at the geometric center of the collision, the space-average fields
weighted by the energy density are smaller in the early stage but
damp slower in the later stage. The space average of squared fields
as well as the EM anomaly $\mathbf{E}\cdot\mathbf{B}$ weighted by
the energy density are also calculated. We give parameterized analytical
formula for these average quantities as functions of time by fitting
numerical results for collisions in the collision energy range $7.7-200$
GeV with different impact parameters.
\end{abstract}
\maketitle

\section{Introduction}

Strong electromagnetic (EM) fields are generated in peripheral heavy-ion
collisions. The dominant component is the magnetic field along the
direction of the orbital angular momentum (OAM) or the reaction plane
(labeled as the $-y$ direction) $B_{y}$ for which a quick estimate
\citep{Kharzeev:2007jp,Asakawa:2010bu} shows that the magnetic field
can reach the order of magnitude of strong interaction (characterized
by the pion mass $m_{\pi}$), $eB_{y}\sim\gamma vZe^{2}/R_{A}^{2}\sim3.5\ m_{\pi}^{2}$
or $10^{18}$ Gauss, in Au+Au collisions at the Relativistic Heavy
Ion Collider (RHIC) at $\sqrt{s_{NN}}=200\text{ GeV}$, and $eB_{y}\sim35\ m_{\pi}^{2}$
or $10^{19}$ Gauss in Pb+Pb collisions at the Large Hadron Collider
(LHC) at $\sqrt{s_{NN}}=2.76\text{ TeV}$. The initial strong magnetic
field might significantly contribute to the initial energy density
and thus is no longer negligible in the plasma evolution \citep{Roy:2015coa}.
This provides a good opportunity for studying the interaction between
the EM fields and the strongly interacting quark/nuclear matter. Several
earlier event-by-event simulations without medium feedback \citep{Skokov:2009qp,Bzdak:2011yy,Voronyuk:2011jd,Deng:2012pc,Bloczynski:2012en}
show that the magnetic field at the geometric center of the collision
reaches its maximum value soon after the collision time and then quickly
decrease towards zero. The magnetic field in the early stage decays
with time as $\sim t^{-3}$ \citep{Hattori:2016emy}, which is mainly
determined by fast-moving spectators. However, once the conductivity
of the matter is considered, the induced Ohm's currents will significantly
slow down the damping of magnetic fields in the later stage, which
has been tested analytically \citep{Tuchin:2013apa,Tuchin:2014iua,Li:2016tel,Chen:2021nxs}
and numerically \citep{McLerran:2013hla,Gursoy:2014aka,Inghirami:2016iru}.
In the case of ideal magnetohydrodynamics, where an infinite electric
conductivity is assumed, the time behavior of magnetic field is estimated
as $\sim t^{-1}$ \citep{Roy:2015kma,Pu:2016ayh,Yan:2021zjc}. Therefore
one can expect some measurable effects because the magnetic field
has enough time to influence the evolution of the hot and dense matter.
For example, the Faraday and Hall effects will result in a charge-odd
directed flow $v_{1}$ \citep{Gursoy:2014aka,Gursoy:2018yai,Inghirami:2019mkc,Oliva:2020mfr,Sun:2021psy},
which has been measured in experiments \citep{Adam:2019wnk,Acharya:2019ijj}.

In recent years, anomalous phenomena driven by magnetic fields have
been widely studied, such as the chiral magnetic effect (CME) \citep{Kharzeev:2007jp,Fukushima:2008xe},
the chiral separation effect (CSE) \citep{Son:2004tq,Metlitski:2005pr,Son:2009tf},
and the chiral magnetic wave (CMW) \citep{Kharzeev:2010gd} (see e.g.
\citep{Huang:2015oca,Kharzeev:2015znc} for reviews). In heavy ion
collisions, the CP symmetry can be spontaneously broken \citep{Kharzeev:2007jp},
which results in an asymmetry between left-handed and right-handed
quarks, described by a nonzero chiral chemical potential $\mu_{5}=(\mu_{R}-\mu_{L})/2$,
where $\mu_{R/L}$ denote the chemical potentials of right-handed
and left-handed quarks respectively. In the CME, the magnetic field
will induce a vector current along its direction, ${\bf j}=[\mu_{5}/(2\pi)]q{\bf B}$,
where $q$ is the quark's electric charge. Tremendous efforts have
been made to search for the CME signal \citep{Abelev:2009ac,Abelev:2009ad,Abelev:2012pa}
in Au+Au or Pb+Pb collisions. The charge separation relative to the
reaction plane has been observed and qualitatively agrees with the
CME prediction. However, large backgrounds from several possible non-CME
contributions \citep{Abelev:2009ac,Adamczyk:2013hsi,Adamczyk:2014mzf,Khachatryan:2016got,Acharya:2017fau,Sirunyan:2017quh}
make it difficult to isolate the CME signal. The isobar collisions
provide a new opportunity for CME search because non-CME contributions
are expected to be identical in Ru+Ru and Zr+Zr collisions, while
the magnetic field in Ru+Ru collisions is about $\sim10\%$ larger
than that in Zr+Zr collisions \citep{Deng:2016knn,Deng:2018dut,Schenke:2019ruo,Shi:2019wzi}.
However, at present no CME signatures have been observed \citep{STAR:2021mii}.
On the other hand, in the CSE, the axial vector current can be induced
along the magnetic field, ${\bf j}_{5}=[\mu_{V}/(2\pi)]q{\bf B}$,
where $\mu_{V}=(\mu_{R}+\mu_{L})/2$ is the vector chemical potential.
The interplay between the CME and the CSE give rise to a collective
wave called the CMW \citep{Kharzeev:2010gd,Burnier:2011bf,Burnier:2012ae,Yee:2013cya}.
In heavy-ion collisions, the CMW is expected to give different elliptic
flows of positive and negative charges \citep{Burnier:2011bf,Ma:2014iva}.
The charge-dependent flows for charged pions have been observed \citep{Adamczyk:2015eqo},
but whether it is the consequence of the CMW is still under debate
because the EM anomaly ${\bf E}\cdot{\bf B}$ can also give similar
effects \citep{Zhao:2019ybo}. All these chiral effects depend on
the magnetic field strength and the matter density in terms of chemical
potentials.

The EM fields produced in heavy ion collisions are highly inhomogeneous
in space-time \citep{Deng:2012pc,Li:2016tel,Zhao:2019crj}. For example,
the magnetic field has a maximum value at the geometric center of
the collision and is much smaller in the edge region. Therefore using
the magnetic field at the geometric center would overestimate the
magnitudes of these chiral effects. In this paper, we propose to calculate
the space-average EM fields weighted by the energy density. This is
based on the fact that the less energy density there is in the region,
the less contribution to the chiral effects it has from the magnetic
field and matter density. We calculate the EM fields by simulations
of Au+Au collisions with the Ultra Relativistic Quantum Molecular
Dynamics (UrQMD) model \citep{Bass:1998ca,Bleicher:1999xi}. Similar
to many other event-by-event simulations, the EM fields generated
by charged particles are given by the Lienard-Wiechert potential in
vacuum. The positions and momenta of charged particles as functions
of time are provided by the simulation using the UrQMD model. Besides
the energy density as the average weight, we also use the charge density
as the weight. We find that the averages weighted by the energy and
charge density make almost no difference in the final results since
the density distributions of the energy and charge are almost the
same in the quark/nuclear matter formed in heavy-ion collisions.

The average EM fields weighted by the charge and energy density can
be applied to estimate the strength of many effects related to EM
fields such as the chiral magnetic effect (CME). In some simulations
of CME through hydrodynamics, the time evolution of the magnetic field
is put by hand instead of self-consistent calculation of fully coupled
fluids and fields. Normally one chooses the magnetic field at one
particular space point such as the geometric center (0,0,0). This
may bring un-controlled errors to the CME signal. A more precise choice
is the average EM fields weighted by the matter density (characterized
by charge or energy) because CME exists in the matter instead of in
vacuum. Note that the CME depends on the axial charge density $n_5$.
In Refs. \citep{Jiang:2016wve,Shi:2017cpu,Hou:2017szz,Lin:2018nxj,Shi:2019wzi,Choudhury:2021jwd},
the initial $n_5$ is set to be proportional to the local entropy density.
It is also reasonable to choose $n_5=\lambda_5 \epsilon$, where $\epsilon$ is the local energy density. This choice also reflects the fact that the gluon topological fluctuations are stronger in matter with higher density. Then one can expect that the charge separation induced by the CME is linear in the energy-density weighted average magnetic field, $\left\langle n_5B\right\rangle=\lambda_5 \left\langle \epsilon B\right\rangle$. Experimental observables for the CME, e.g., the three-point $\gamma$ correlator \citep{Voloshin:2004vk} and the $\delta$ correlator \citep{Bzdak:2012ia}, are quadratically proportional to the charge separation \citep{Shi:2019wzi,Choudhury:2021jwd}. Therefore they are quadratically proportional to the energy-density weighted average magnetic field. On the other hand,
the average squared EM fields are of special importance to
estimate the strength of the vector mesons' spin alignment, see Refs.
\citep{Sheng:2019kmk,Sheng:2020ghv}.

The paper is organized as follows. In Sec. \ref{sec:weight-average}
we define the space-average EM fields weighted by the energy or charge
density. We give the formula for Lienard-Wiechert potentials of EM
fields used in later simulations using the UrQMD model. Here we assume
that one nucleus moves along $+z$ direction with its center located
at $x=b/2$ and the other nucleus moves along $-z$ direction with
its center located at $x=-b/2$, where $b$ is the impact parameter.
So the OAM or the reaction plane is in $-y$ direction. The results
for the space-average fields are presented in Sec. \ref{sec:Transverse-magnetic-field}.
Only the $y$-component of the average magnetic field, $\left\langle eB_{y}\right\rangle _{E}$
(the index $E$ means the energy as the weight), is nonzero, while
other components, $\left\langle eB_{x}\right\rangle _{E}$ and $\left\langle eB_{z}\right\rangle _{E}$,
as well as $\left\langle eE_{i}\right\rangle _{E}$ $(i=x,y,z)$ are
all vanishing due to the symmetry of the collision. Then the results
for $\left\langle eB_{y}\right\rangle _{E}$ are presented for different
collision energies and impact parameters. A comparison of the space-average
fields with those at the geometric center of the collision has been
made. In Sec. \ref{sec:Squared-field} and Sec. \ref{sec:EM-anomaly}
we present the results for the space-averages of squared fields $(eB_{i})^{2}$
and $(eE_{i})^{2}$ as well as the EM anomaly $e^{2}{\bf E}\cdot{\bf B}$,
respectively. The parameterized analytical formula for $\left\langle eB_{y}\right\rangle _{E}$,
$\left\langle (eB_{i})^{2}\right\rangle _{E}$ and $\left\langle (eE_{i})^{2}\right\rangle _{E}$
with $i=x,y,z$, and $\left\langle e^{2}{\bf E}\cdot{\bf B}\right\rangle _{E}$
are given in Sec. \ref{sec:Analytical-formula-of} by fitting numerical
results for collisions at energies in the range $7.7-200$ GeV with
different impact parameters. A summary and an outlook are given in
Sec. \ref{sec:Summary-and-conclusions}.

\section{Space-average EM fields weighted by energy and charge density \label{sec:weight-average}}

The time evolution and spatial distributions of EM fields in heavy-ion
collisions have been extensively investigated \citep{Deng:2012pc,Bloczynski:2012en,Tuchin:2013apa,Zhao:2019crj}.
Most of studies focus on the time evolution of fields at the geometric
center $(x,y,z)=(0,0,0)$ or spatial distributions at some specific
time. However, since EM fields vary in both space and time, their
overall effects on physical observables should be at the average level
in the full volume and lifetime of the quark/nuclear matter. Considering
the fact that the matter and EM fields are coupled, to quantify the
average effects of EM fields, we define the space-average fields weighted
by the energy and charge density
\begin{eqnarray}
\left\langle \mathbf{F}\right\rangle _{E}(t) & \equiv & \frac{\int d^{3}\mathbf{r}\varepsilon(t,\mathbf{r})\mathbf{F}(t,\mathbf{r})}{\int d^{3}\mathbf{r}\varepsilon(t,\mathbf{r})},\nonumber \\
\left\langle \mathbf{F}\right\rangle _{C}(t) & \equiv & \frac{\int d^{3}\mathbf{r}\rho(t,\mathbf{r})\mathbf{F}(t,\mathbf{r})}{\int d^{3}\mathbf{x}\rho(t,\mathbf{r})},\label{eq:average-fields}
\end{eqnarray}
where $\mathbf{F}$ represents the electric field $\mathbf{E}$ or
the magnetic field $\mathbf{B}$ as functions of space-time, $\rho(t,\mathbf{r})$
is the (net) charge density, $\varepsilon(t,\mathbf{r})$ is the energy
density, both as functions of space-time, and the indices 'C' and
'E' label the energy and charge density respectively. In the numerical
calculations, the integral over space costs a lot of computing time,
so we divide the whole space into grids, and the integrals in Eq.
(\ref{eq:average-fields}) are converted to sums over grids as
\begin{eqnarray}
\left\langle \mathbf{F}\right\rangle _{E}(t) & \equiv & \frac{\sum_{i}\varepsilon_{i}(t)\mathbf{F}_{i}(t)}{\sum_{i}\varepsilon_{i}(t)},\nonumber \\
\left\langle \mathbf{F}\right\rangle _{C}(t) & \equiv & \frac{\sum_{i}\rho_{i}(t)\mathbf{F}_{i}(t)}{\sum_{i}\rho_{i}(t)},\label{eq:weighted-average}
\end{eqnarray}
where $\rho_{i}(t)$ and $\varepsilon_{i}(t)$ are the net charge
and energy in the $i$-th grid at the time $t$, respectively, and
$\mathbf{F}_{i}(t)$ denotes the EM field at the center of the same
grid. When evaluating the charge or energy density in each grid, we
only consider particles in the mid-rapidity range $-0.5<Y<0.5$ in
the fireball ($Y$ denotes the momentum rapidity). When calculating
EM fields, however, all charged particles including those in spectators
are taken into account.

We use the Lienard-Wiechert potential to calculate the EM fields as
functions of space-time
\begin{eqnarray}
\mathbf{B}(t,\mathbf{r}) & = & \frac{1}{4\pi}\sum_{n}q_{n}\frac{\mathbf{R}_{n}-R_{n}{\bf v}_{n}}{(R_{n}-\mathbf{R}_{n}\cdot\mathbf{v}_{n})^{3}}(1-v_{n}^{2})\Theta_{n}(t_{n}^{\text{ret}}),\nonumber \\
\mathbf{E}(t,\mathbf{r}) & = & \frac{1}{4\pi}\sum_{n}q_{n}\frac{{\bf v}_{n}\times\mathbf{R}_{n}}{(R_{n}-\mathbf{R}_{n}\cdot\mathbf{v}_{n})^{3}}(1-v_{n}^{2})\Theta_{n}(t_{n}^{\text{ret}}),\label{eq:LW-potential}
\end{eqnarray}
where $n$ labels the charged particle, and $\mathbf{R}_{n}\equiv{\bf r}-\mathbf{r}_{n}(t_{n}^{\text{ret}})$
with $R_{n}=\left|\mathbf{R}_{n}\right|$ and $\mathbf{r}_{n}(t_{n}^{\text{ret}})$
being the location of the $n$-th particle at the retarded time $t_{n}^{\text{ret}}=t-\left|\mathbf{r}-\mathbf{r}_{n}(t_{n}^{\text{ret}})\right|$.
If the $n$-th particle does not exist at the retarded time $t_{n}^{\text{ret}}$,
i.e., if it is created after $t_{n}^{\text{ret}}$ or annihilated
before $t_{n}^{\text{ret}}$, then it will not contribute to the EM
field at $(t,\mathbf{r})$. In Eq. (\ref{eq:LW-potential}), we have
introduced a step function $\Theta_{n}(t)$ to describe the particle's
lifetime,
\begin{equation}
\Theta_{n}(t)=\left\{ \begin{array}{cc}
1, & t_{n}^{\text{create}}<t<t_{n}^{\text{annihilate}}\\
0, & \text{else}
\end{array}\right.
\end{equation}
where $t_{n}^{\text{create}}$ and $t_{n}^{\text{annihilate}}$ are
the creation time and annihilation time of the $n$-th particle. The
positions and momenta of charged particles at any time are given by
UrQMD simulations.

\section{Average fields \label{sec:Transverse-magnetic-field}}

In this section, we present the calculations of the space-average
fields weighted by the energy density by Eq. (\ref{eq:weighted-average}).
Due to the symmetry of the collision, the only non-vanishing component
is $\left\langle B_{y}\right\rangle $ in non-central collisions,
while all other components $\left\langle B_{x}\right\rangle $, $\left\langle B_{z}\right\rangle $,
and $\left\langle E_{i}\right\rangle $ $(i=x,y,z)$, are vanishing.
So we only focus on $\left\langle B_{y}\right\rangle $ in this section.

\subsection{Spatial distribution}

\begin{figure}[tbh]
\begin{centering}
\includegraphics[width=16cm]{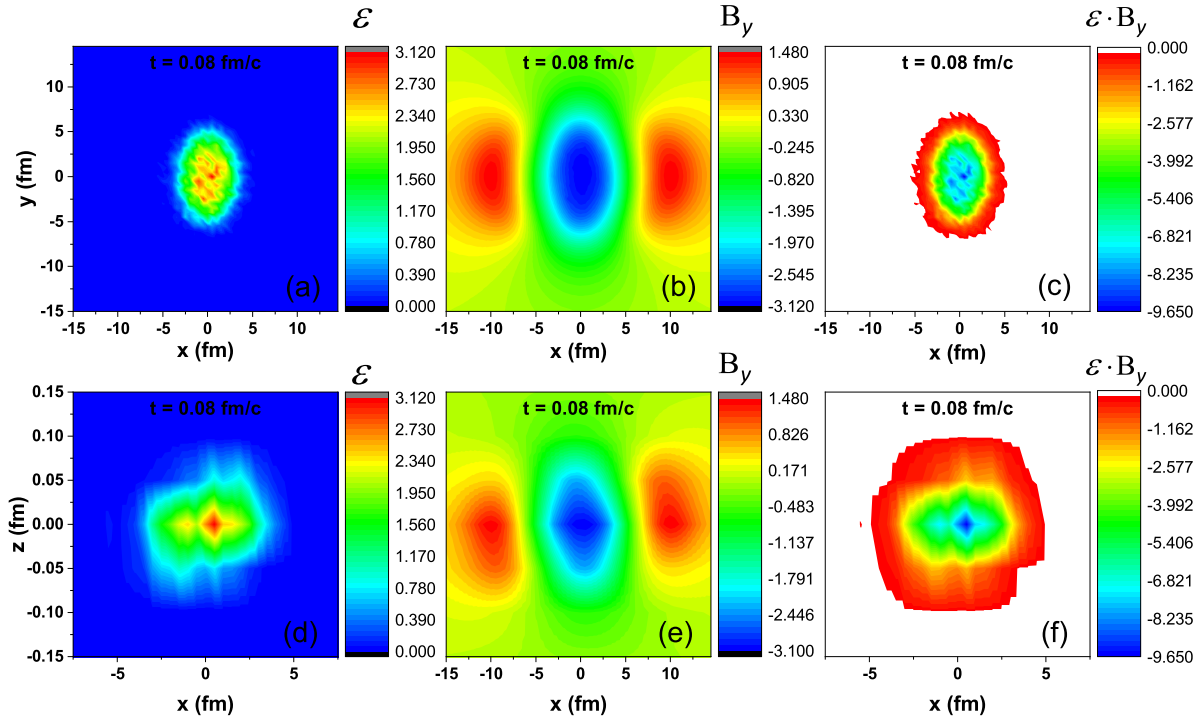}
\par\end{centering}
\caption{\label{fig:eDenProfiles}The spatial distributions of the energy density
[(a) and (c)], the magnetic field $eB_{y}$ [(b) and (d)], and the
product of the energy density and $eB_{y}$ [(e) and (f)] in the transverse plane
[(a), (b), and (c)] and reaction plane [(d), (e), and (f)] at $t=0.08$ fm/c in Au+Au
collisions at 200 GeV and $b=9$ fm.}
\end{figure}

The spatial distributions of the energy density and the magnetic field
are shown in Fig. \ref{fig:eDenProfiles} for Au+Au collisions at
200 GeV at $t=0.08$ fm/c. Figures \ref{fig:eDenProfiles}(a) and (d) shows the
energy densities in the transverse and reaction plane, respectively.
As we have mentioned, when calculating the energy density, we only
counts particles in the mid-rapidity range $-0.5<Y<0.5$. So the influence
from spectators and boundary region of the quark/nuclear matter is
eliminated. Spatial distributions of $B_{y}$ are shown in Figs. \ref{fig:eDenProfiles}(b) and (e). One can see that $B_{y}$ in the central region is negative
while it is positive in the peripheral region: the spatial distribution
of the magnetic field looks like that of a magnet with its north pole
pointing to $-y$ direction. Figures \ref{fig:eDenProfiles}(c) and (f) shows the distribution
of $\varepsilon B_{y}$, one can see that only in the central region
is $\varepsilon B_{y}$ non-vanishing.

\subsection{Grid-size dependence}

The expression in Eq. (\ref{eq:weighted-average}) involves summations
over grids, in which we use the field value at the center of each
grid as the mean value in that grid. The EM fields produced in heavy-ion
collisions are space-time dependent, thus the size of the grid should
be small enough to achieve a reasonable precision. On the other hand,
the computing time increases dramatically with the decrease of the
grid size. So we have to find an appropriate grid size to balance
these two contradictory constraints. In this subsection, we study
the grid-size dependence of $\left\langle eB_{y}\right\rangle _{E}$
in order to find an optimized value for the grid size.

\begin{figure}[tb]
\begin{centering}
\includegraphics[width=8cm]{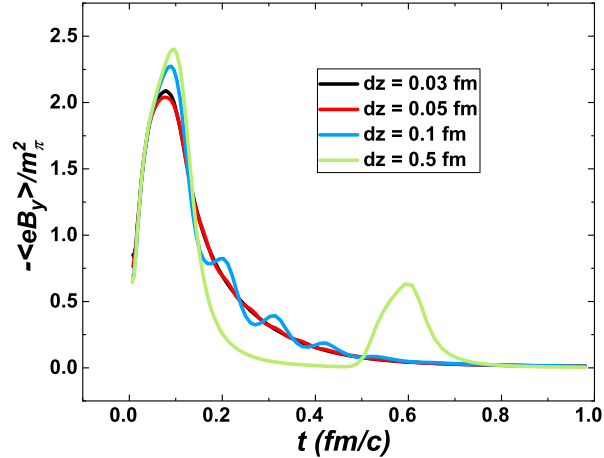}
\par\end{centering}
\caption{\label{fig:Grid-size-dependence}The time dependence of $\left\langle eB_{y}\right\rangle _{E}$
on the longitudinal grid size in Au+Au collisions at 200 GeV and $b=7$
fm.}
\end{figure}

We consider Au+Au collisions at 200 GeV and $b=7$ fm. When calculating
the average value, we choose the space volume as $-15\text{ fm}<\text{\ensuremath{x}, \ensuremath{y}}<15\text{ fm}$
and $-10\text{ fm}<z<10\text{ fm}$, and divide it into grids with
the grid size $dx$, $dy$, and $dz$. Figure \ref{fig:Grid-size-dependence}
shows $\left\langle eB_{y}\right\rangle _{E}$ in the unit $m_{\pi}^{2}$
as functions of time with $dx=dy=0.5\text{ fm}$ and various values
of $dz$. One can see that there are peaks when $dz=0.5$ fm and $0.1$
fm. This is because the typical length scale of the magnetic field's
variation is smaller than the grid size in the longitudinal direction
due to the Lorentz contraction. In this case, the magnetic field at
the grid center cannot represent its mean value in the grid. The peaks
in magnetic fields arise when spectators, which generate a narrow
distribution of the magnetic field in the $z$ direction, are close
to centers of some grids. We notice that results become smooth enough
for $dz=0.05$ fm (red line) and $dz=0.03$ fm (black line). Therefore,
we will choose $dz=0.05$ fm in later calculations, which is small
enough to obtain smooth magnetic fields.

We also study the dependence of $\left\langle eB_{y}\right\rangle _{E}$
on the grid size in the transverse plane. We fix $dz=0.05$ fm and
take $dx=dy=0.05,0.1,0.5$ fm, respectively. We see that the values
of $\left\langle eB_{y}\right\rangle _{E}$ are almost independent
of $dx$ and $dy$ because the magnetic field slowly varies in the
transverse direction. In later calculations we will choose $dx=dy=0.5$
fm.

\subsection{Impact parameter and collision energy dependences}

\begin{figure}[tbh]
\begin{centering}
\includegraphics[width=8cm]{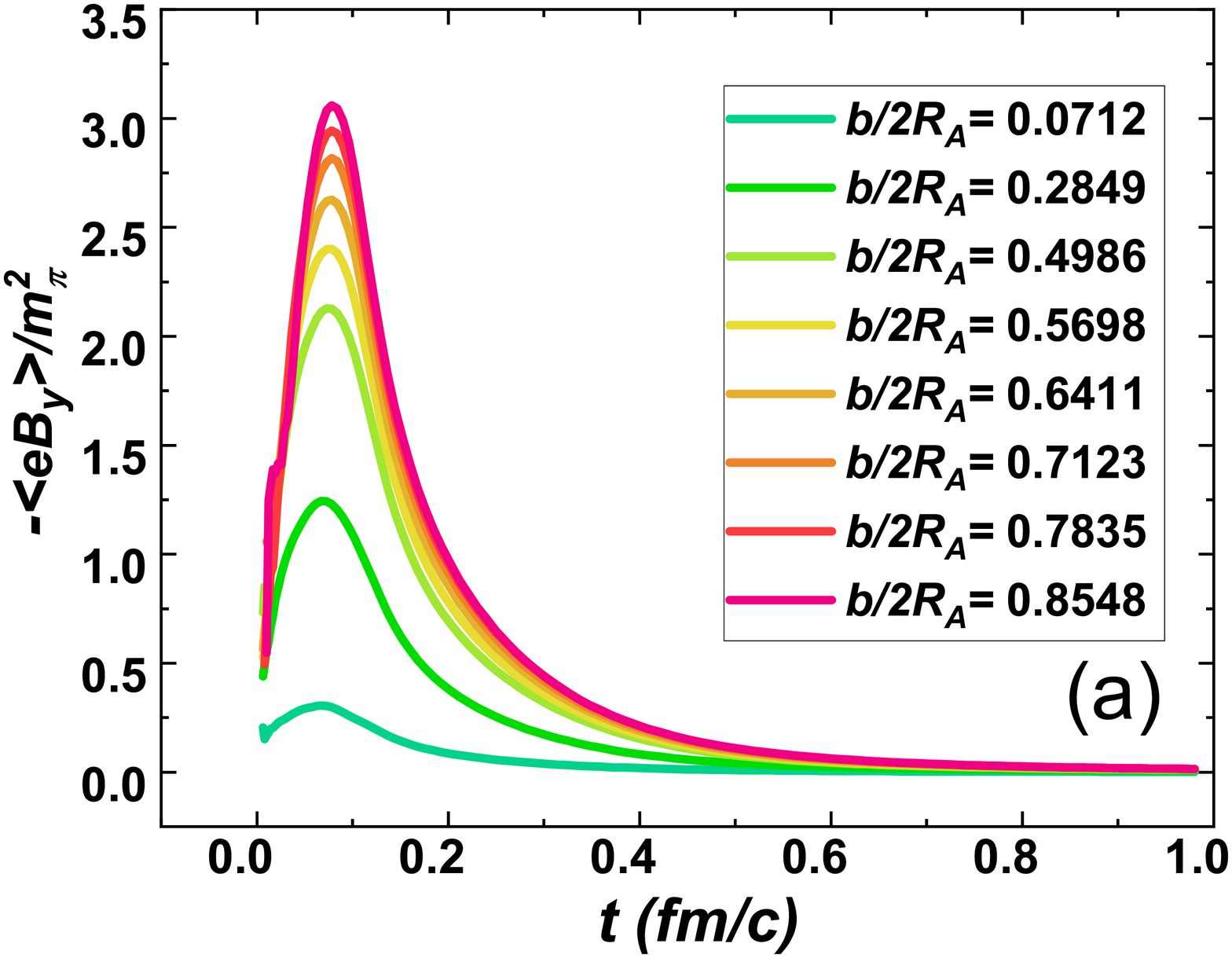}\includegraphics[width=8cm]{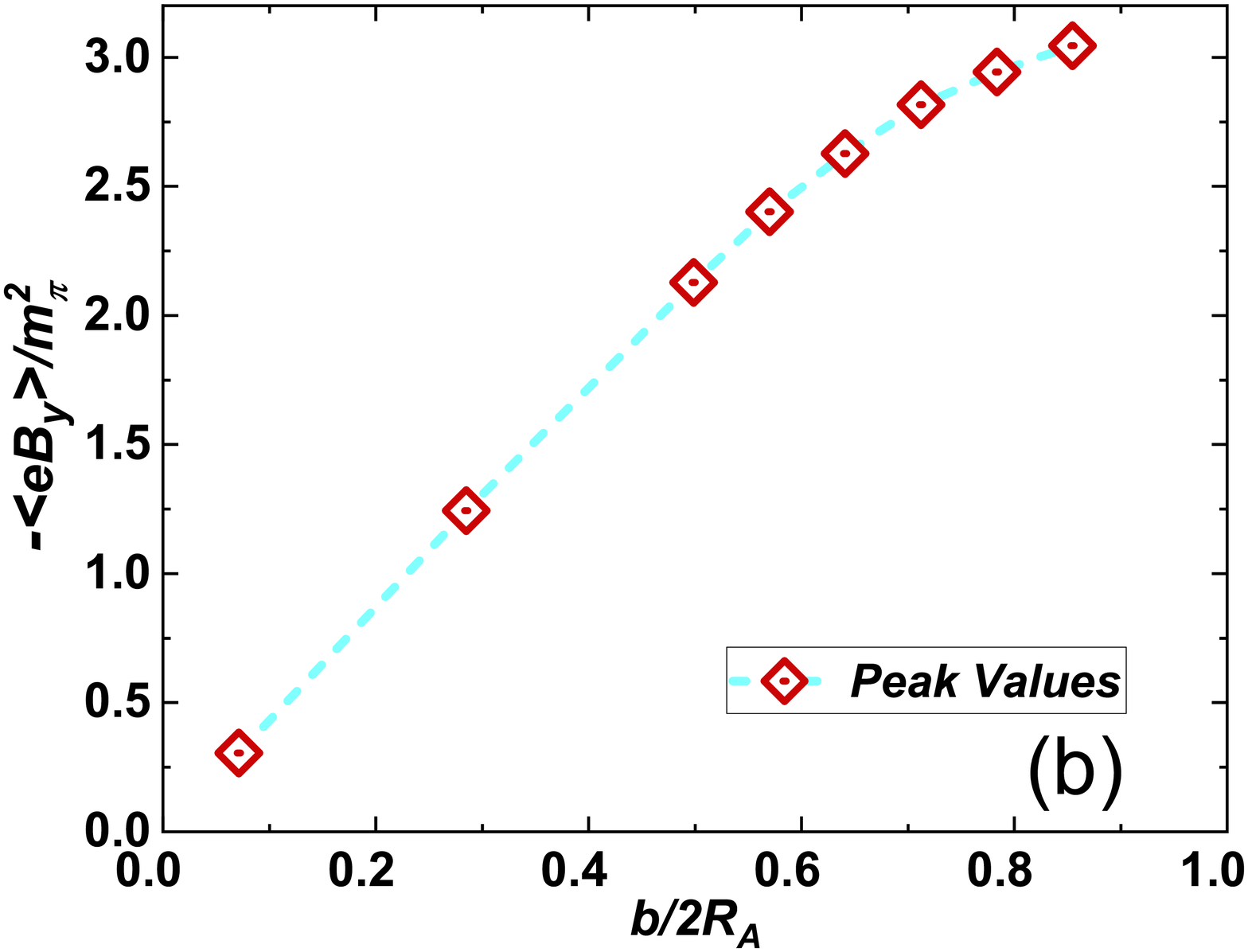}
\par\end{centering}
\caption{\label{fig:MagneticField-ImpactParameter}(a): the time evolution
of $\left\langle eB_{y}\right\rangle _{E}$ for various impact parameters
in Au+Au collisions at 200 GeV. (b): the peak value of $\left\langle eB_{y}\right\rangle _{E}$
as a function of the impact parameter.}
\end{figure}

The impact parameter dependence of $\left\langle eB_{y}\right\rangle _{E}$
is shown in Fig. \ref{fig:MagneticField-ImpactParameter} for Au+Au
collisions at 200 GeV and $b=$1, 4, 7, 8, 9, 10, 11, 12 fm. We see
in Fig. \ref{fig:MagneticField-ImpactParameter}(a) that all $\left\langle eB_{y}\right\rangle _{E}$
have peak values at about $t=0.08$ fm/c after the collision, and
then fastly falls to the values 2 or 3 orders of magnitudes smaller
than the peak values in about 1 fm/c. We plot the peak values as functions
of the impact parameter in Fig. \ref{fig:MagneticField-ImpactParameter}(b).
We observe that $\left\langle eB_{y}\right\rangle _{E}$ is proportional
to $b$ for small $b$, similar to the behavior of $B_{y}$ at one
specific space-time point $(t,\mathbf{x})=(0,0,0,0)$ in Ref. \citep{Deng:2012pc}.

\begin{figure}[tbh]
\begin{centering}
\includegraphics[width=8cm]{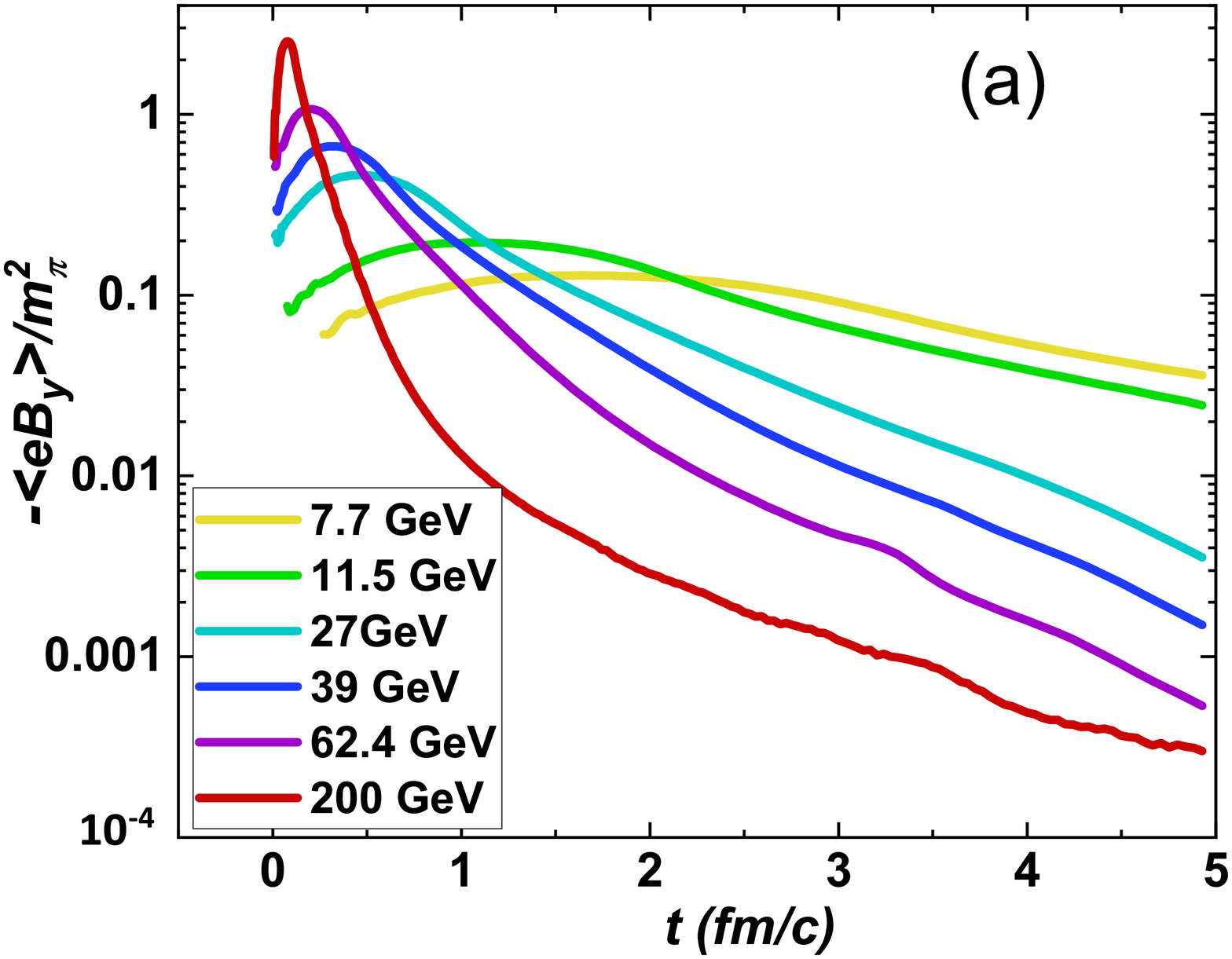}\includegraphics[width=8cm]{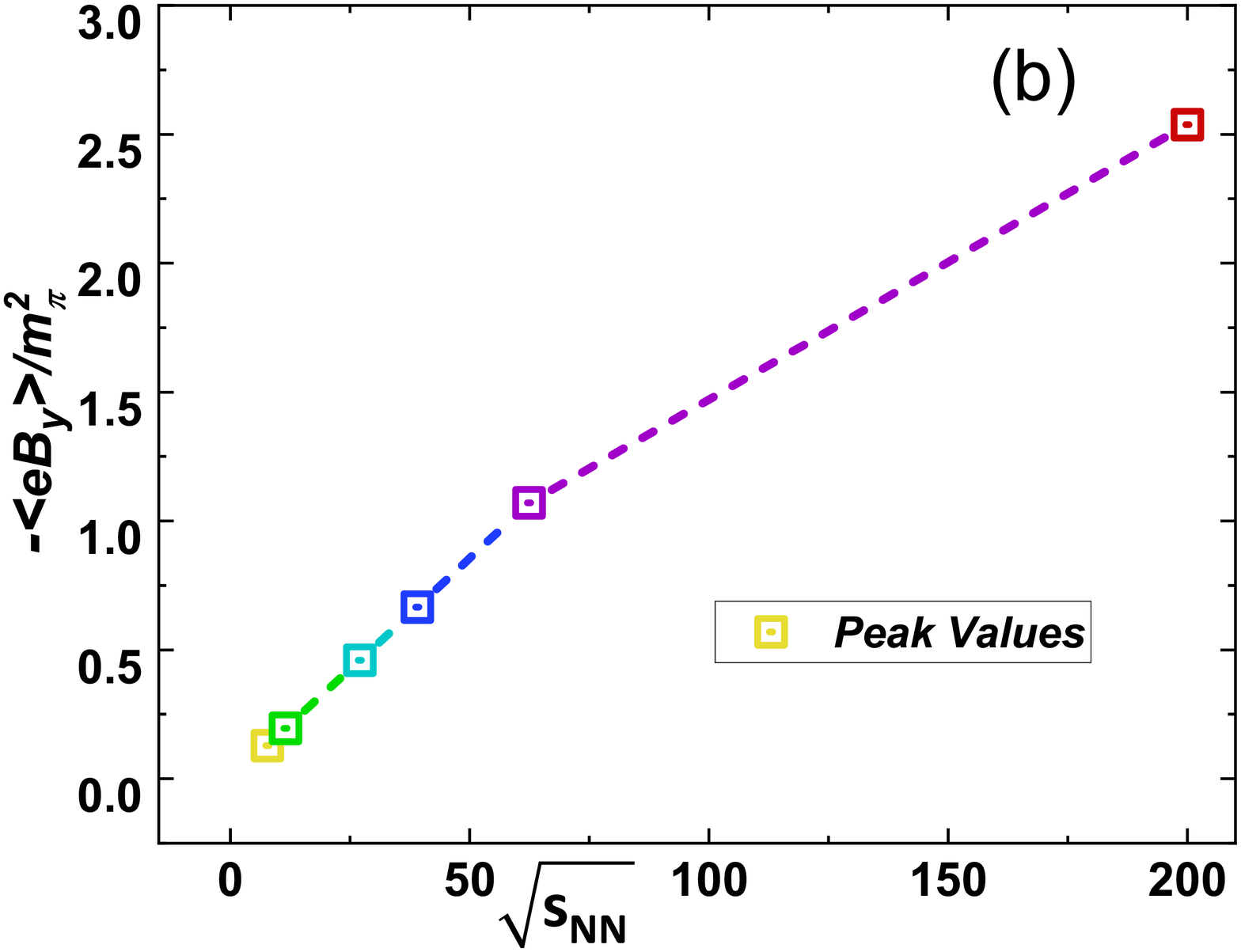}
\par\end{centering}
\caption{\label{fig:CollisionEnergyDependence}(a): the time evolution
of the space-average magnetic field weighted by the energy density,
$\left\langle eB_{y}\right\rangle _{E}$, in Au+Au collision at several
collision energies and $b=9$ fm. (b): the peak values of
$\left\langle eB_{y}\right\rangle _{E}$ as a function of the collision
energy. }
\end{figure}

The time evolution of $\left\langle eB_{y}\right\rangle _{E}$ at
different collision energies and $b=9$ fm is shown in Fig. \ref{fig:CollisionEnergyDependence}(a).
The maximum values of $\left\langle eB_{y}\right\rangle _{E}$ are
almost proportional to the collision energy as shown in Fig. \ref{fig:CollisionEnergyDependence}(b), similar to the
behavior of $B_{y}$ at one specific space-time $(t,\mathbf{r})=(0,0,0,0)$
in Ref. \citep{Deng:2012pc}. We also observe that $\left\langle eB_{y}\right\rangle _{E}$
reach maximum values earlier at higher than lower collision energies.
Meanwhile, $\left\langle eB_{y}\right\rangle _{E}$ decrease slower
or live longer at lower collision energies. This is because the magnetic
field is mainly generated by spectators moving with the velocity proportional
to the collision energy. At very high collision energies, spectators
of two nuclei go through each other in such a short time that makes
$B_{y}$ behave like a pulse.

\subsection{Comparison with fields at geometric center}

\begin{figure}[tb]
\begin{centering}
\includegraphics[width=8cm]{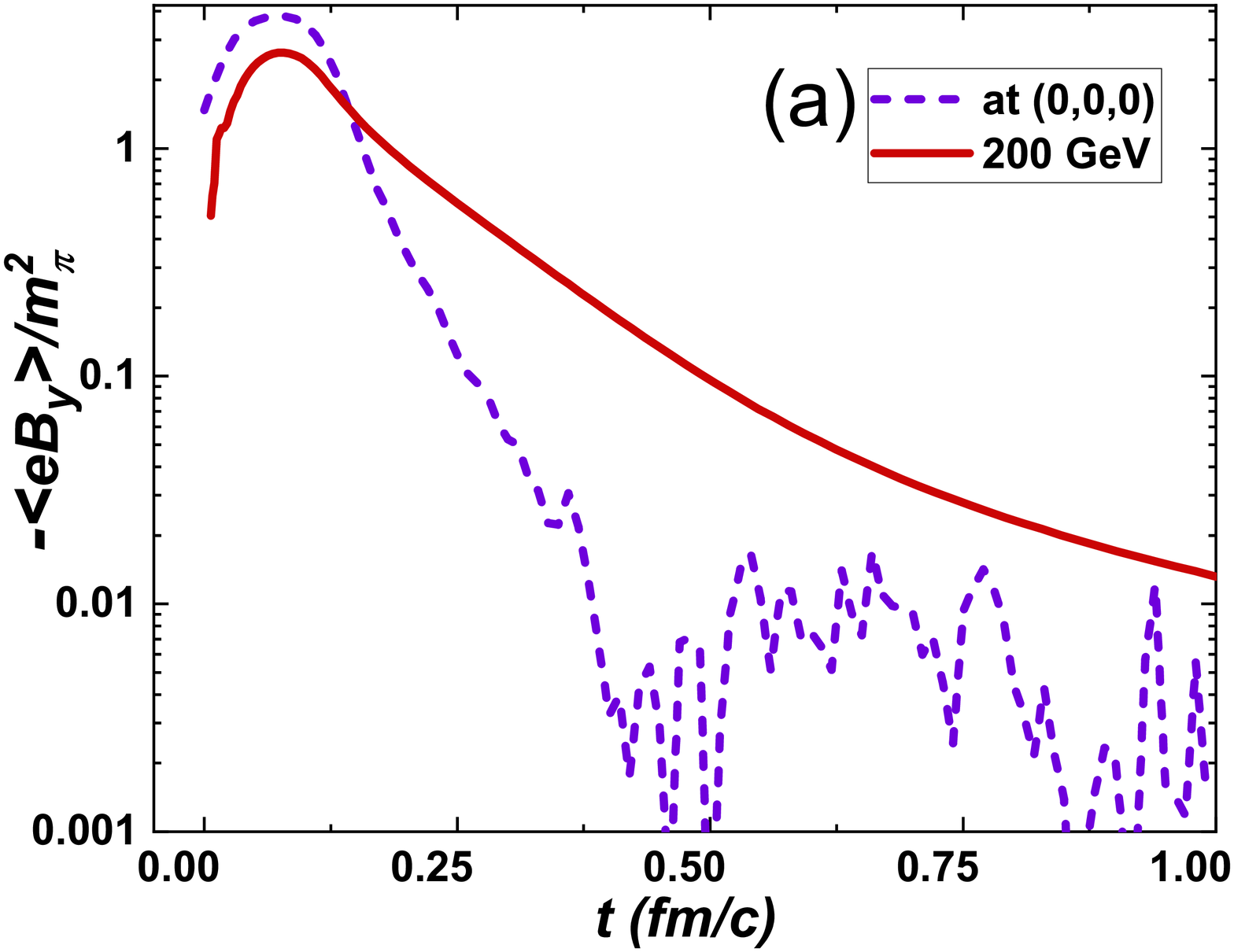}\includegraphics[width=8cm]{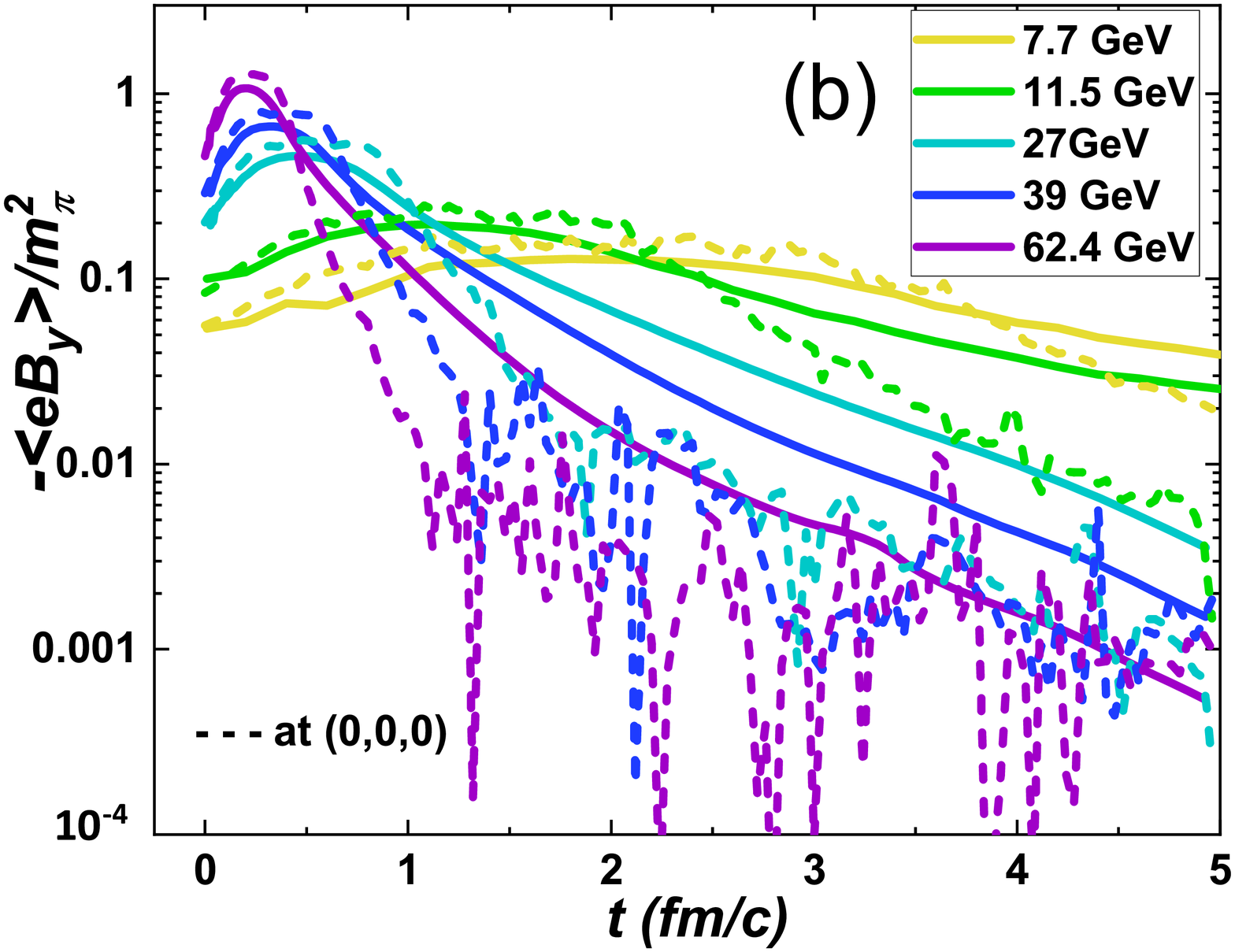}
\par\end{centering}
\caption{\label{fig:CompareWithOrigin}The time evolution of $\left\langle eB_{y}\right\rangle _{E}$
and $eB_{y}$ at the geometric center $\mathbf{r}=(0,0,0)$ at 200
GeV [(a)] and lower energies [(b)] and $b=9$ fm in
Au+Au collisions.}
\end{figure}

In Fig. \ref{fig:CompareWithOrigin} we make a comparison of $\left\langle eB_{y}\right\rangle _{E}$
with $eB_{y}$ at the space point $\mathbf{r}=(0,0,0)$ or the geometric
center as functions of time {[}denoted as $eB_{y}(t,0,0,0)${]} for
collisions at 200 GeV [Fig. \ref{fig:CompareWithOrigin}(a)] and lower energies [Fig. \ref{fig:CompareWithOrigin}(b)]
and $b=9$ fm. We notice that the peak values of $\left\langle eB_{y}\right\rangle _{E}$
(solid lines) are much smaller, fall much slower or live longer than
$eB_{y}(t,0,0,0)$ at all collision energies. This is because the
fireball is expanding and regions close to spectators have larger
$B_{y}$ than at the geometric center. The much longer lives of $\left\langle eB_{y}\right\rangle _{E}$
than $eB_{y}(t,0,0,0)$ show that it is more appropriate and accurate
to use the average field in calculations of any field related effects
than the field at a particular space-time point such as the geometric
center.

\subsection{Comparison between energy and charge density weight}

\begin{figure}[tbh]
\begin{centering}
\includegraphics[width=8cm]{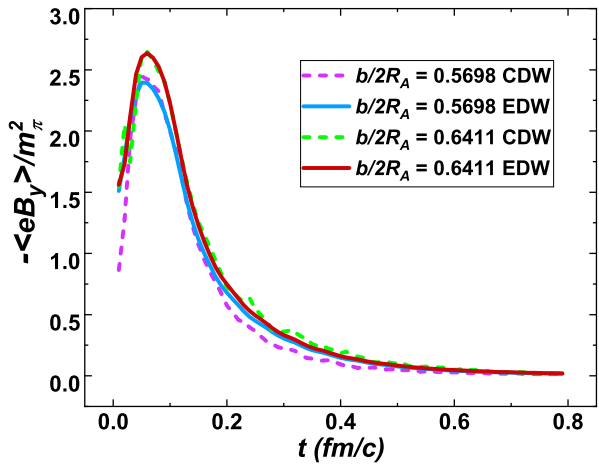}
\par\end{centering}
\caption{\label{fig:Energy-charge-density-comp}Comparison of $\left\langle eB_{y}\right\rangle _{E}$
(solid lines) and $\left\langle eB_{y}\right\rangle _{C}$ (dashed
lines) as functions of time in Au+Au collisions at 200 GeV and $b=8,9$
fm.}
\end{figure}

As shown in Eq. (\ref{eq:weighted-average}), one can calculate space-average
fields weighted either by the energy or charge density. In Fig. \ref{fig:Energy-charge-density-comp},
we make a comparison of average fields with two weights in Au+Au collision
at 200 GeV and $b=8,9$ fm. We see that the results of $\left\langle eB_{y}\right\rangle _{E}$
(solid lines) are smoother than those of $\left\langle eB_{y}\right\rangle _{C}$
(dashed lines). If we take averages over sufficiently large number
of events, fluctuations in average fields weighted by the charge density
are expected to be suppressed.

\section{Squared fields \label{sec:Squared-field}}

\begin{figure}[tbh]
\begin{centering}
\includegraphics[width=5.5cm]{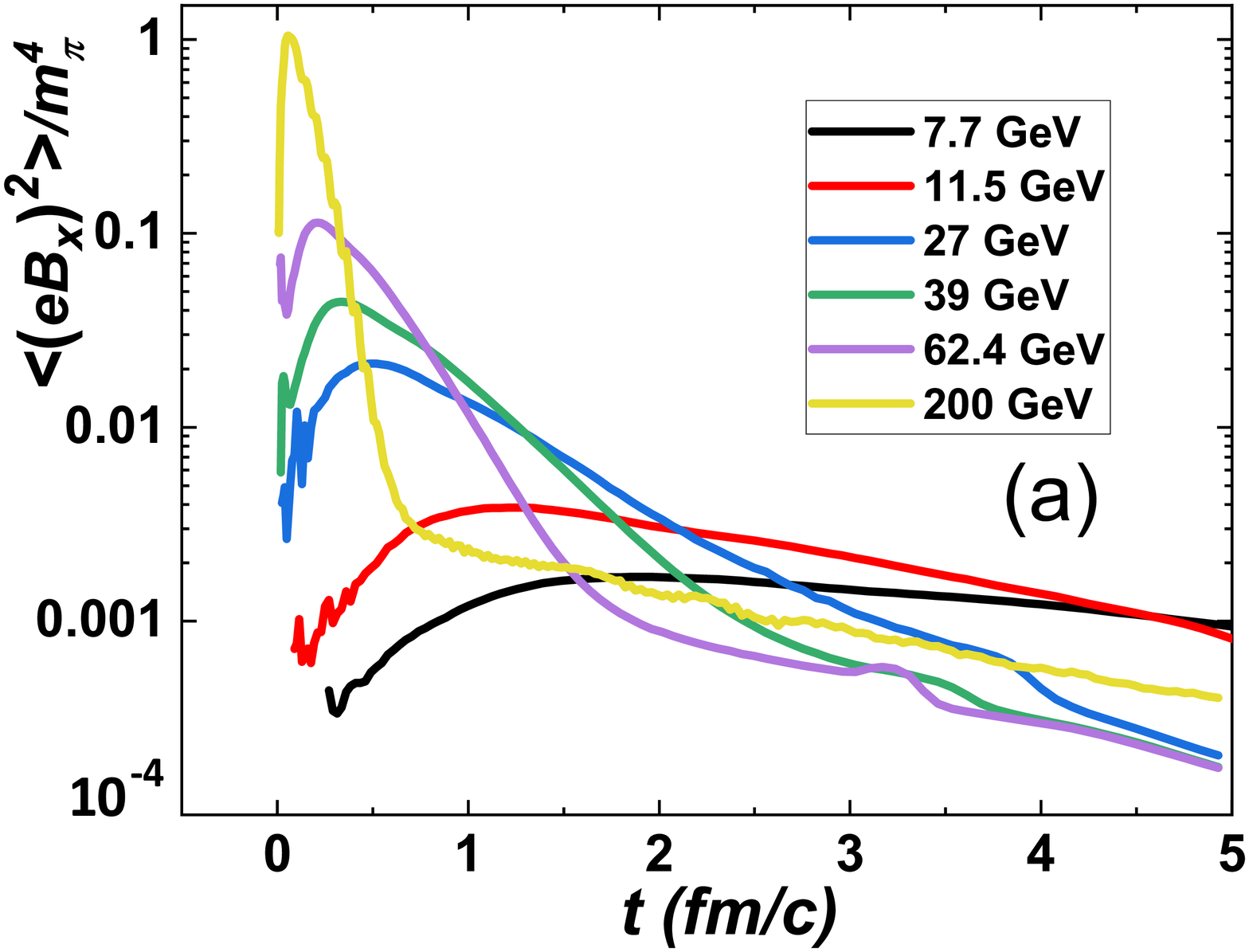}\includegraphics[width=5.5cm]{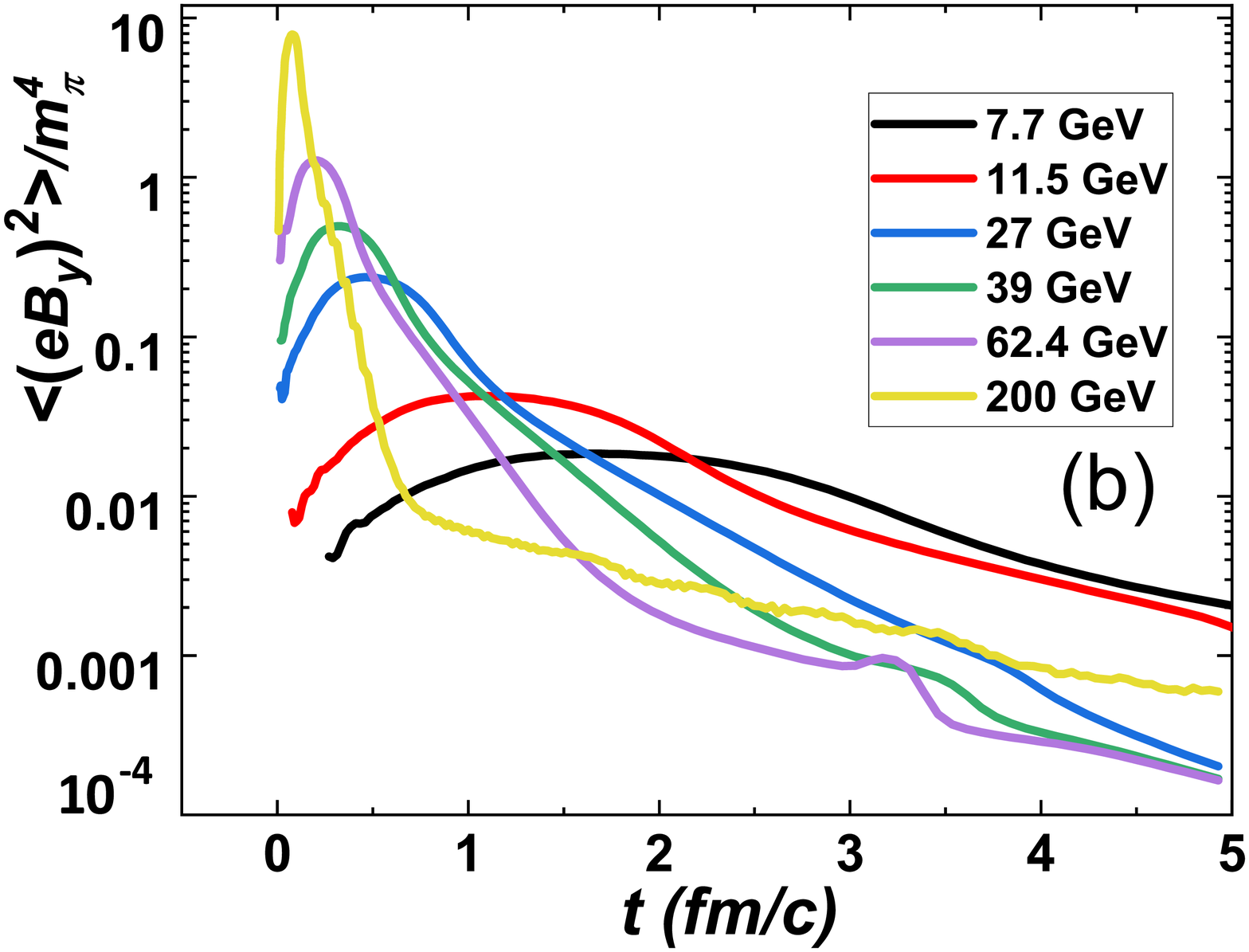}\includegraphics[width=5.5cm]{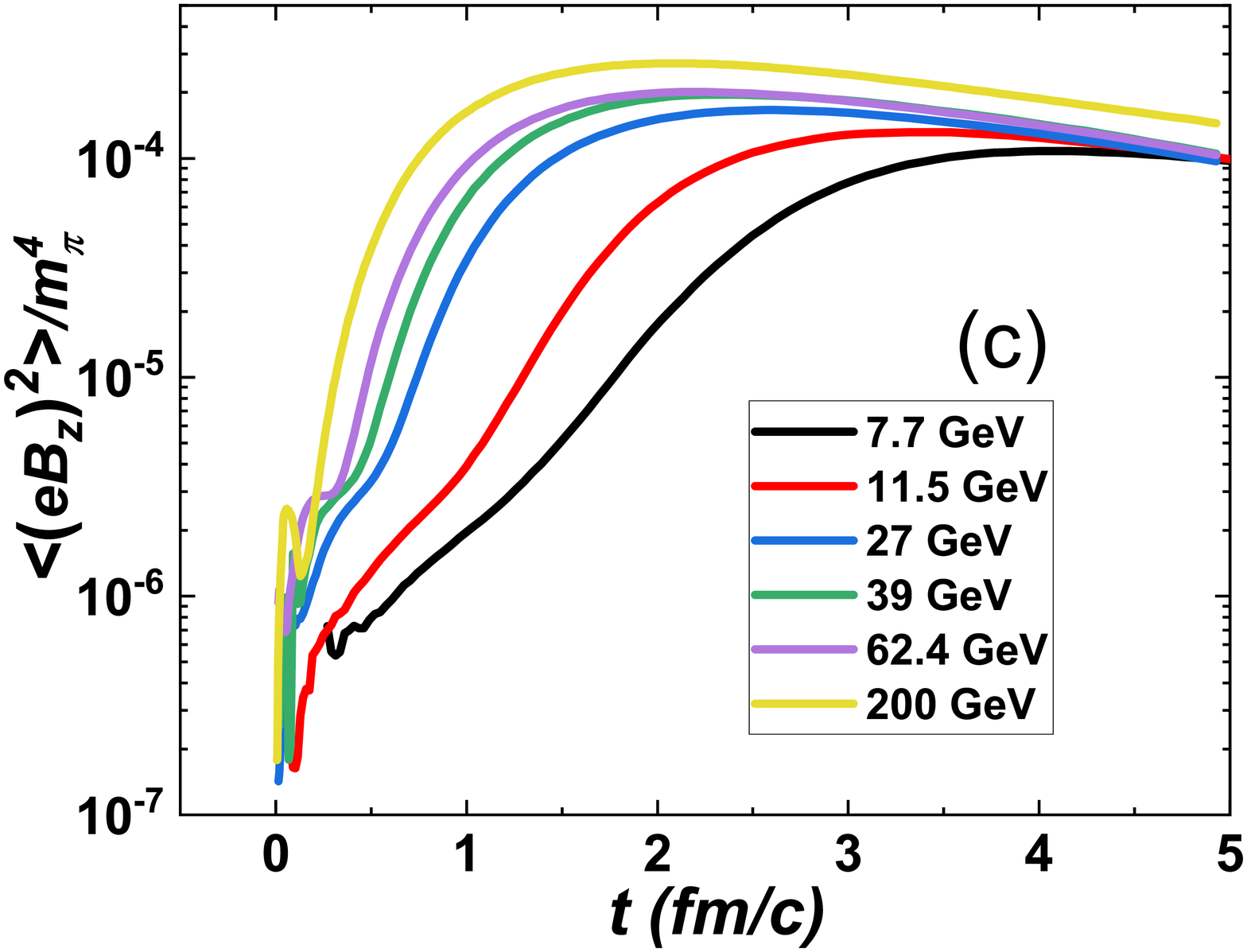}
\par\end{centering}
\caption{\label{fig:SqauredFields}The time evolution of $\left\langle (eB_{i})^{2}\right\rangle {}_{E}$
with $i=x,y,z$ at collision energies ranging from 7.7 GeV to 200
GeV in Au+Au collisions and $b=9$ fm.}
\end{figure}

\begin{figure}[tbh]
\begin{centering}
\includegraphics[width=5.5cm]{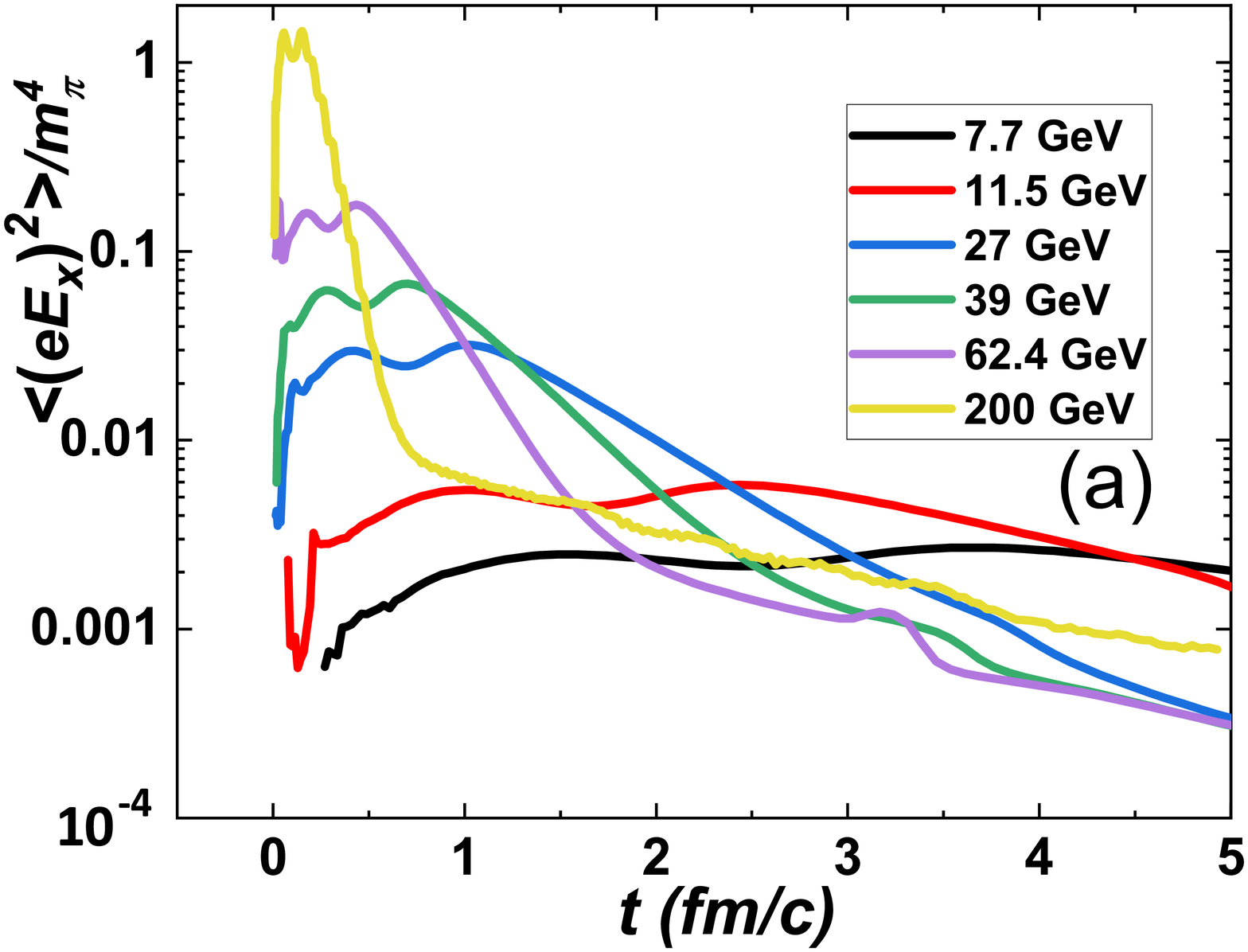}\includegraphics[width=5.5cm]{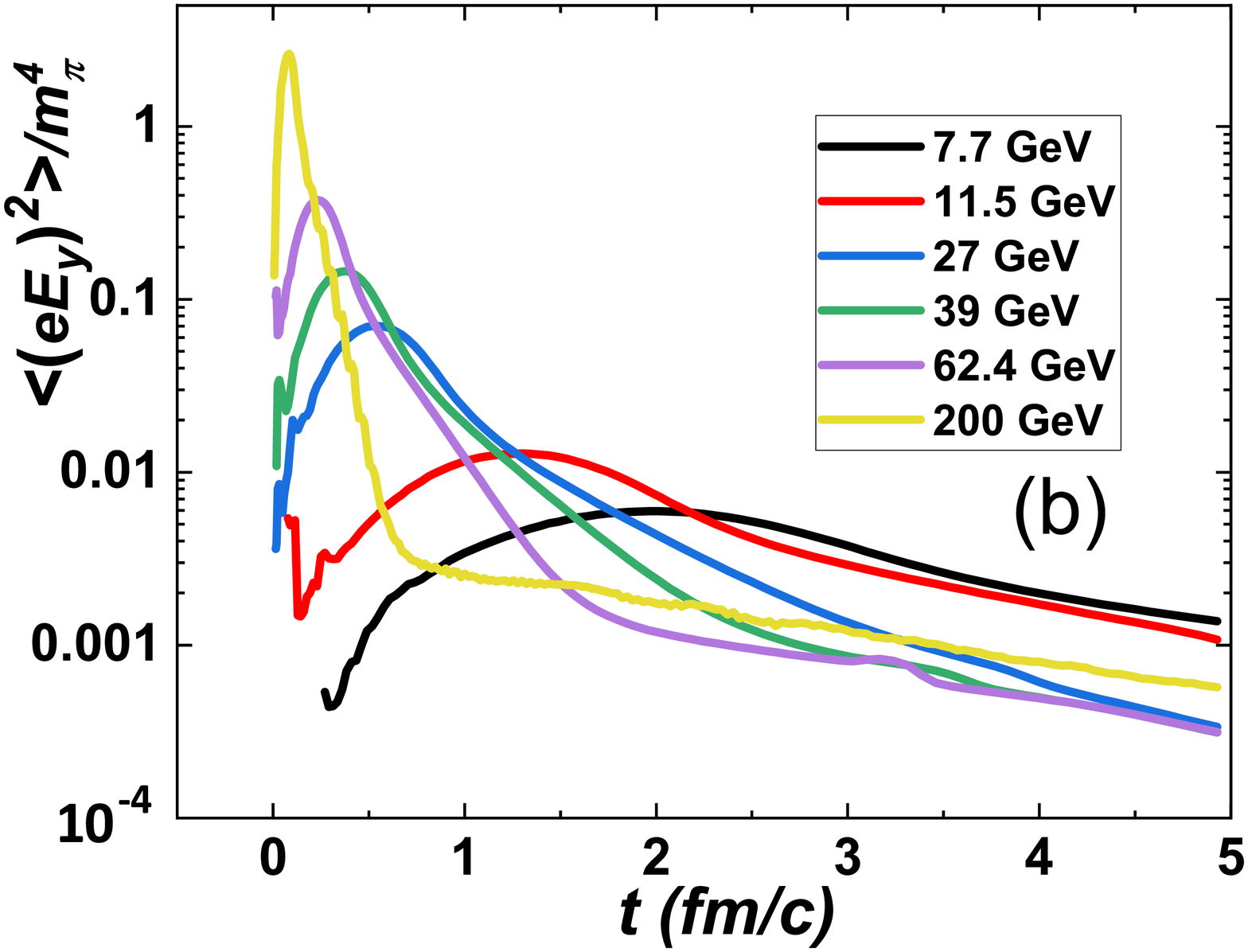}\includegraphics[width=5.5cm]{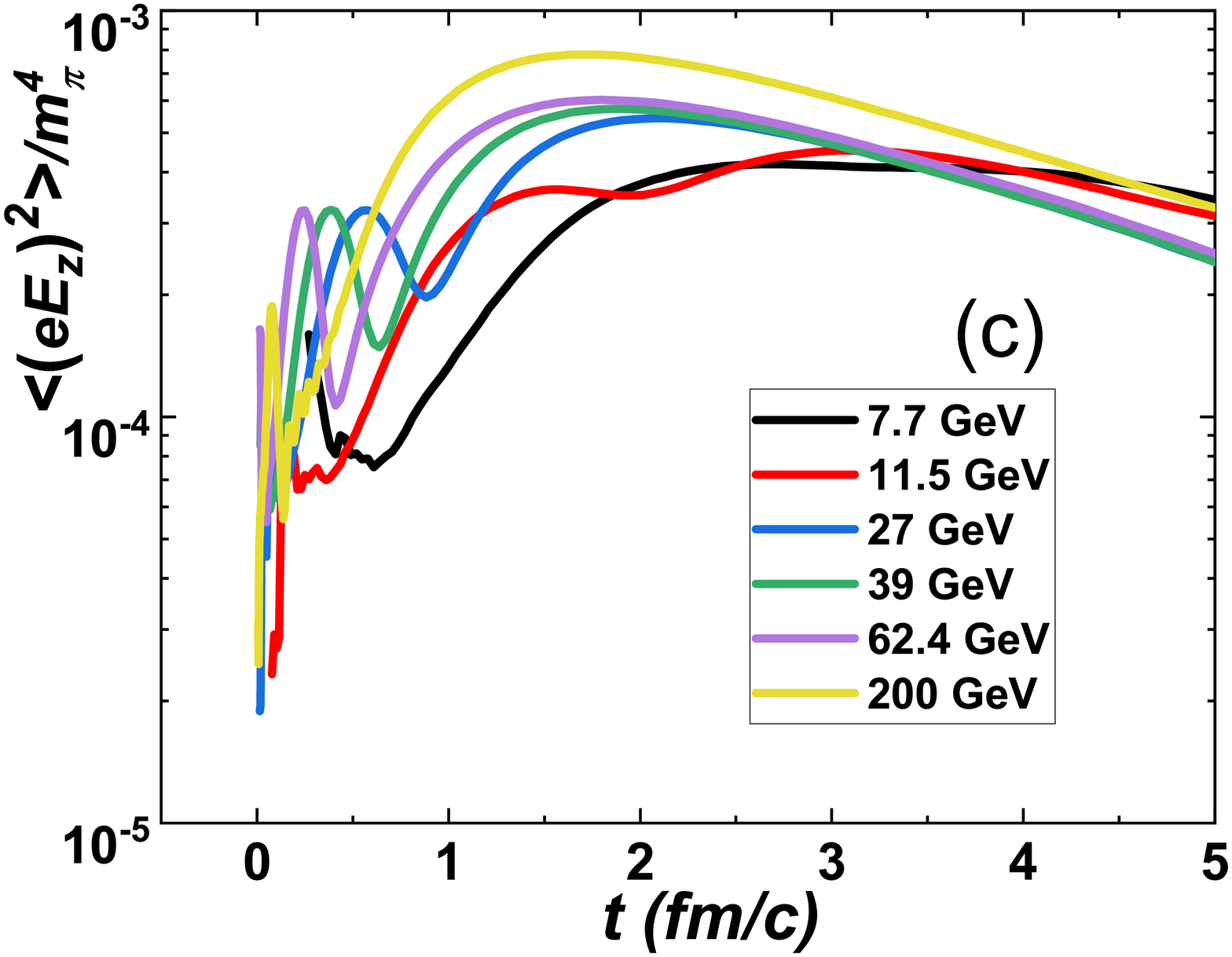}
\par\end{centering}
\caption{\label{fig:SquaredElectricfields}The time evolution of $\left\langle (eE_{i})^{2}\right\rangle {}_{E}$
with $i=x,y,z$ at collision energies ranging from 7.7 GeV to 200
GeV in Au+Au collisions and $b=9$ fm.}
\end{figure}

In this section we calculate the space averages of squared electric
and magnetic fields in Au+Au collisions at energies ranging from 7.7
GeV to 200 GeV and $b=9\text{ fm}$ in the central rapidity region.
The averages of squared electric and magnetic fields play important
roles in the spin alignment of vector mesons \citep{Liang:2004xn,Sheng:2019kmk,Sheng:2020ghv}.
The results for $\left\langle (eB_{i})^{2}\right\rangle {}_{E}$ with
$i=x,y,z$ are shown in Fig. \ref{fig:SqauredFields} and those for
$\left\langle (eE_{i})^{2}\right\rangle {}_{E}$ are shown in Fig.
\ref{fig:SquaredElectricfields}. We see in Fig. \ref{fig:SqauredFields}
that at the same collision energy, the peak value of $\left\langle (eB_{y})^{2}\right\rangle {}_{E}$
is about one order of magnitude larger than that of $\left\langle (eB_{x})^{2}\right\rangle {}_{E}$
and about two (lower energies) to four (higher energies) orders of
magnitude larger than that of $\left\langle (eB_{z})^{2}\right\rangle {}_{E}$.
For electric fields, as shown in Fig. \ref{fig:SquaredElectricfields},
at the same collision energy, the peak value of $\left\langle (eE_{x})^{2}\right\rangle {}_{E}$
is in the same order of magnitude as that of $\left\langle (eE_{y})^{2}\right\rangle {}_{E}$,
both are about one (lower energies) to three (higher energies) orders
of magnitude larger than that of $\left\langle (eE_{z})^{2}\right\rangle {}_{E}$.

\begin{figure}[tbh]
\begin{centering}
\includegraphics[width=5.5cm]{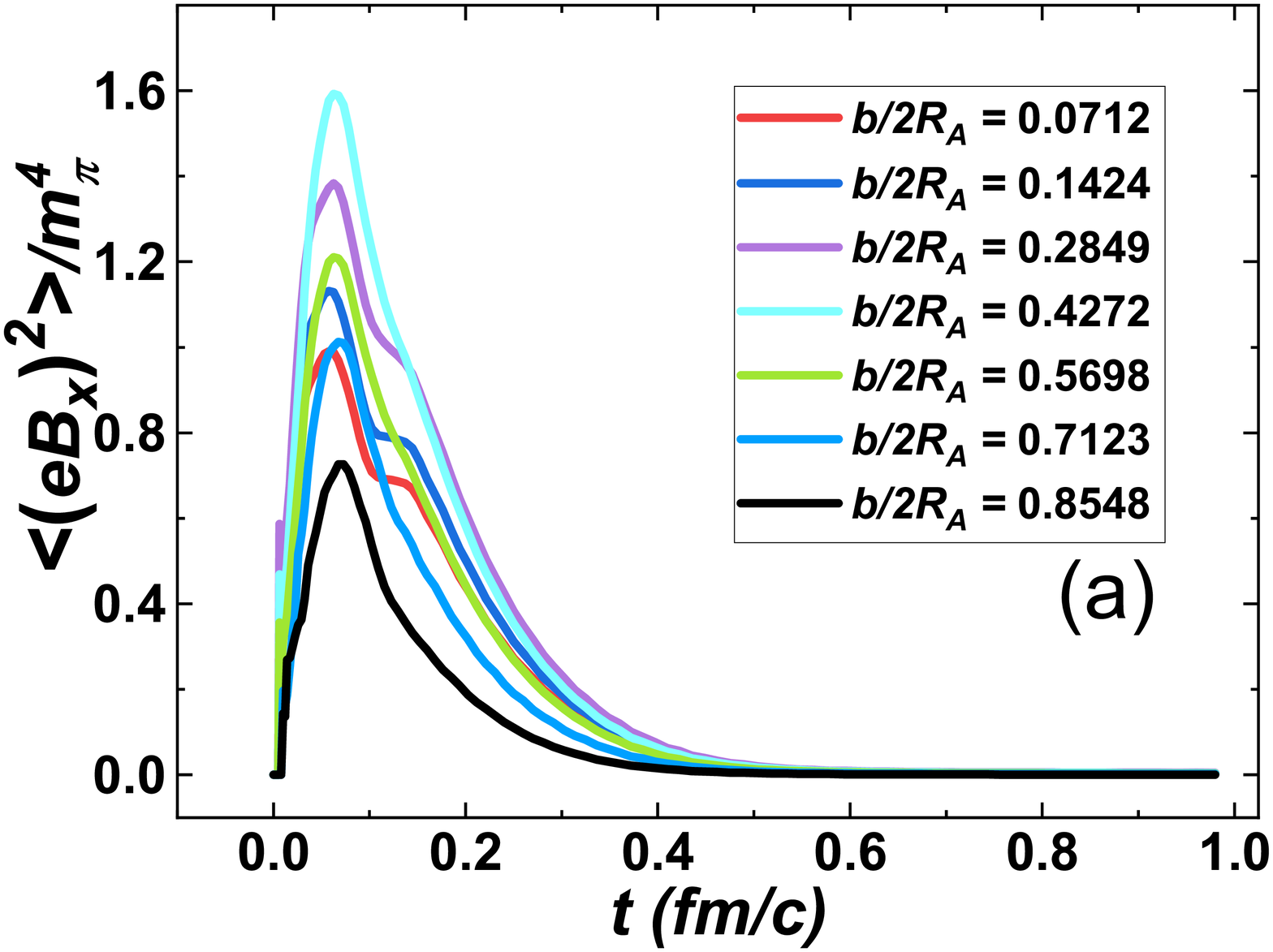}\includegraphics[width=5.5cm]{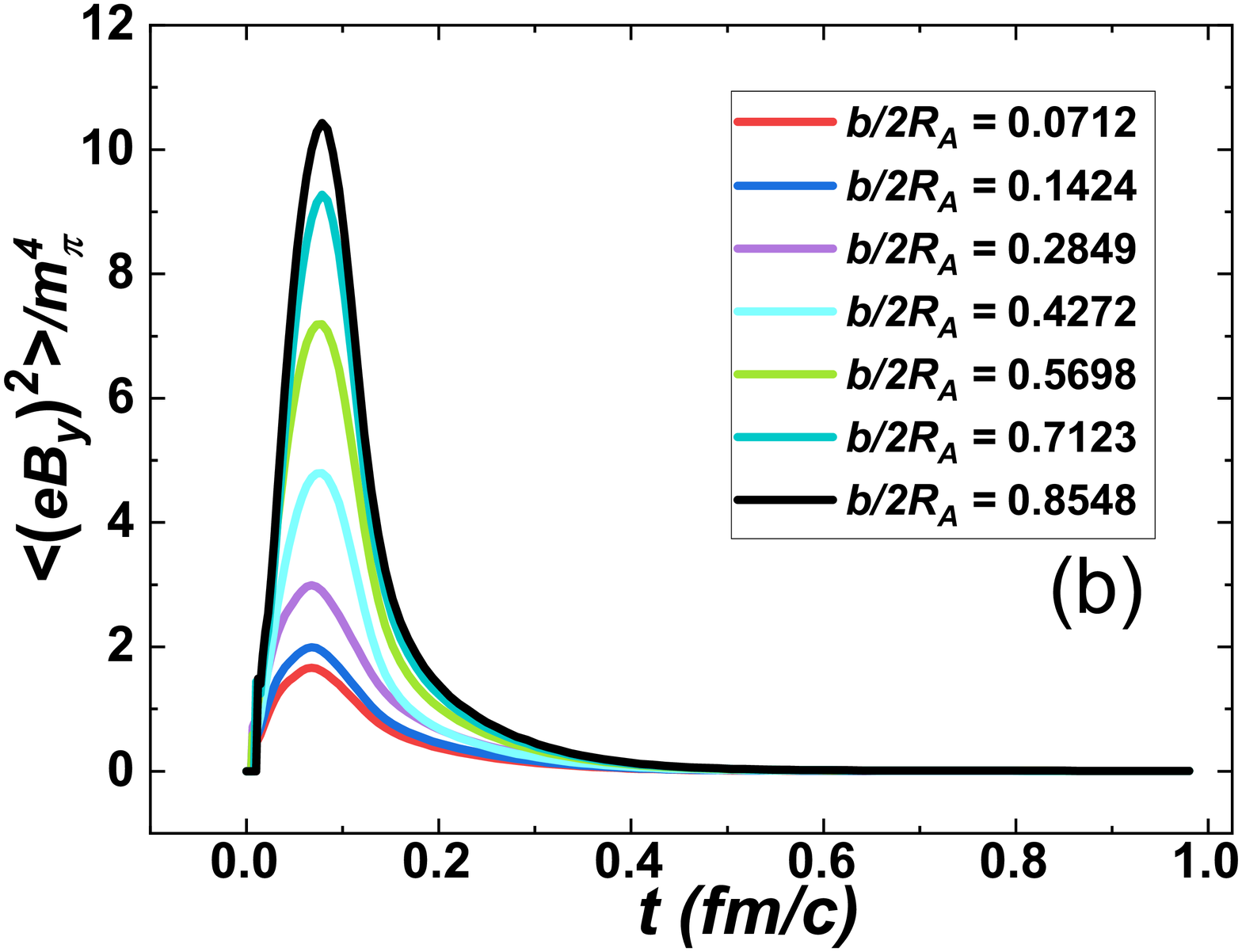}\includegraphics[width=5.5cm]{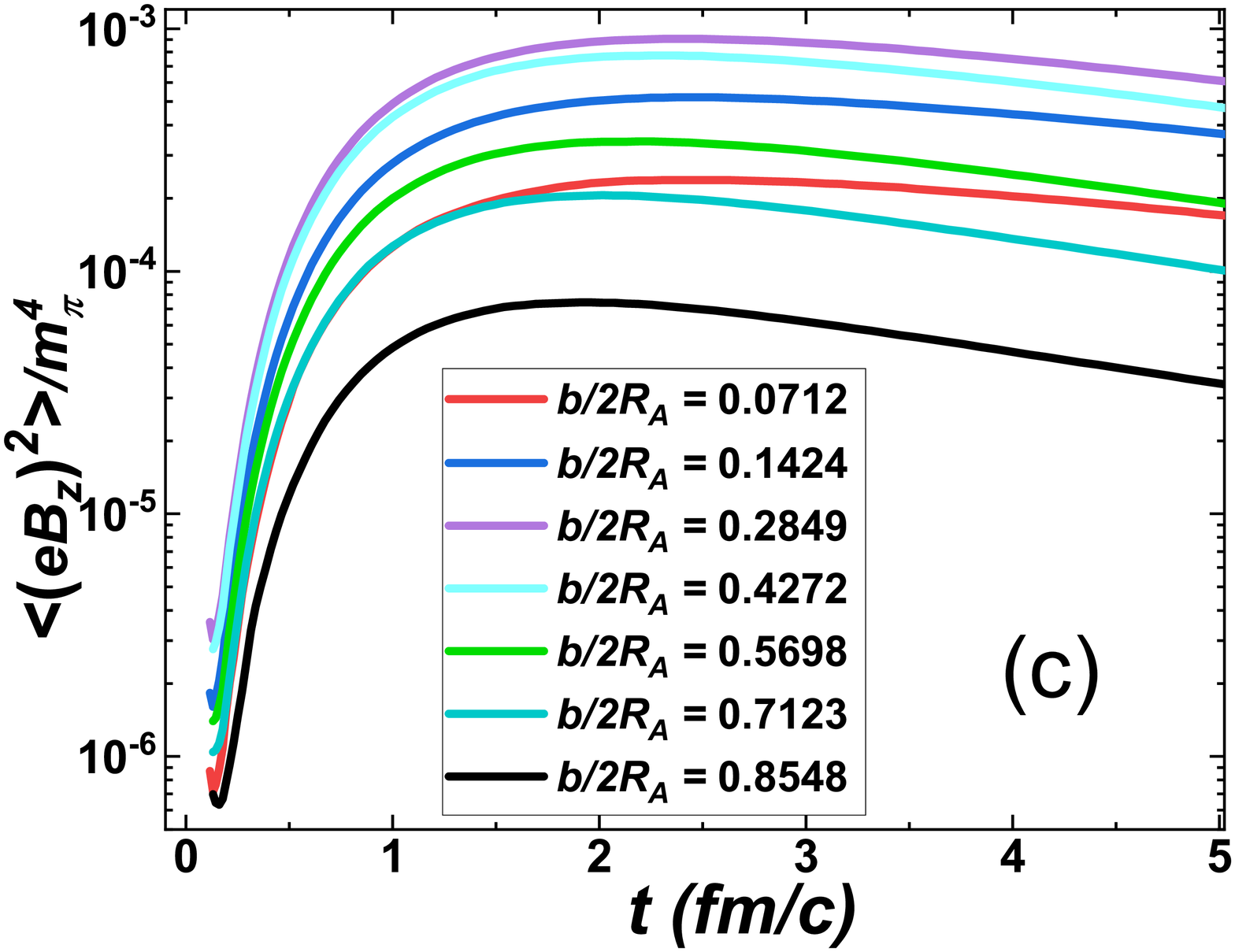}
\par\end{centering}
\caption{\label{fig:SqauredFields_imp}Time evolution of $\left\langle (eB_{i})^{2}\right\rangle _{E}$
with $i=x,y,z$ for various impact parameters in Au+Au collisions
at $200$ GeV.}
\end{figure}

\begin{figure}[tbh]
\begin{centering}
\includegraphics[width=5.5cm]{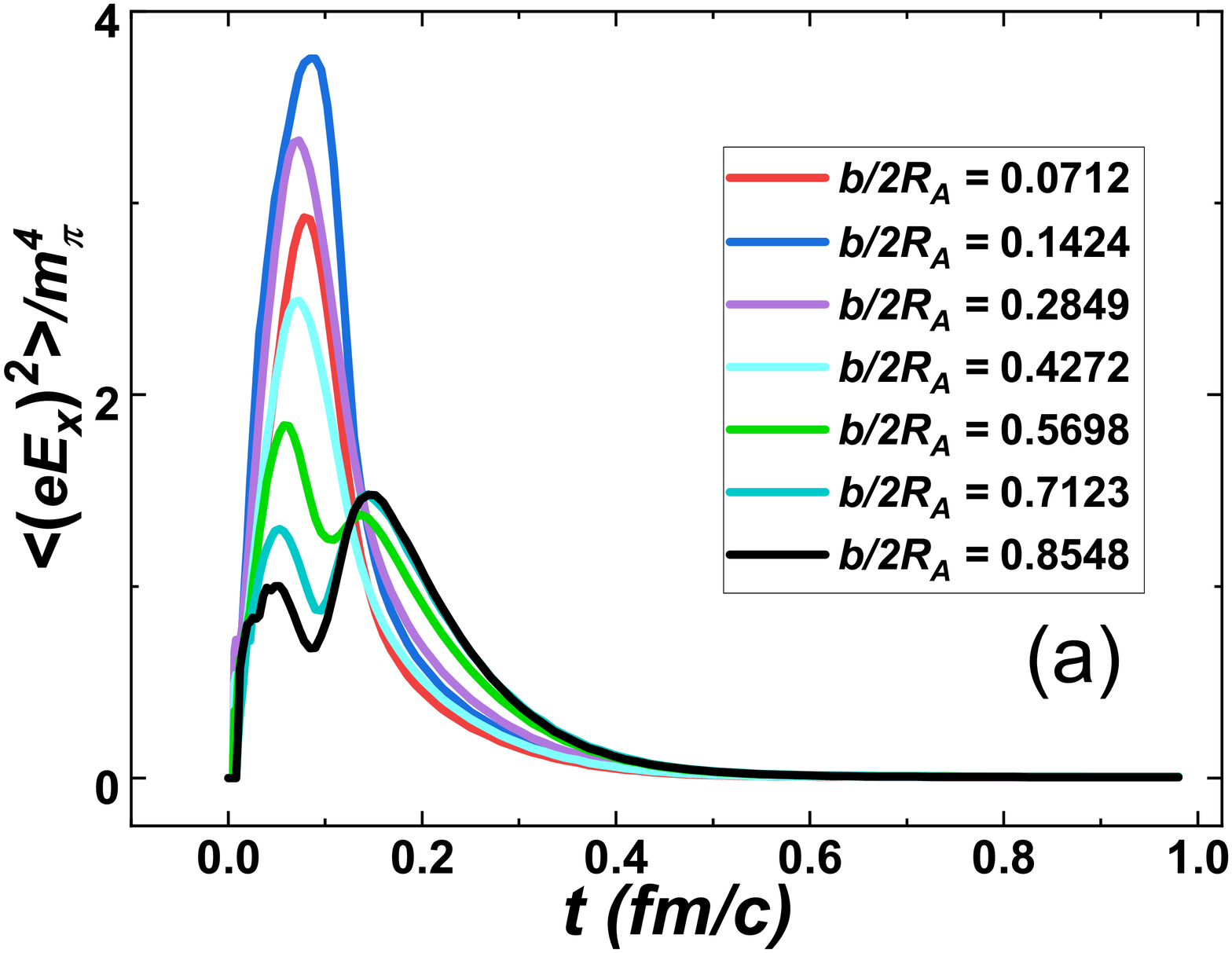}\includegraphics[width=5.5cm]{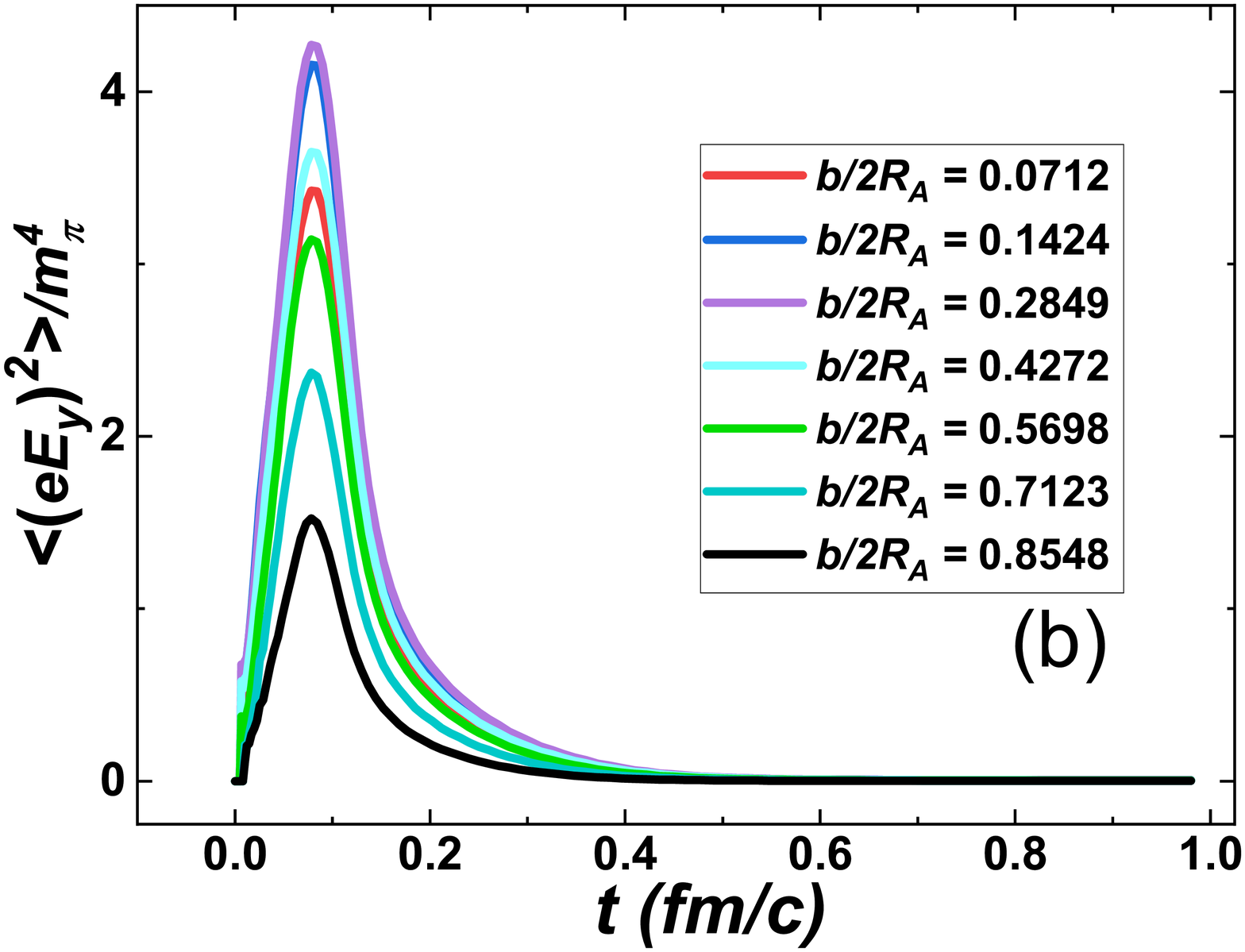}\includegraphics[width=5.5cm]{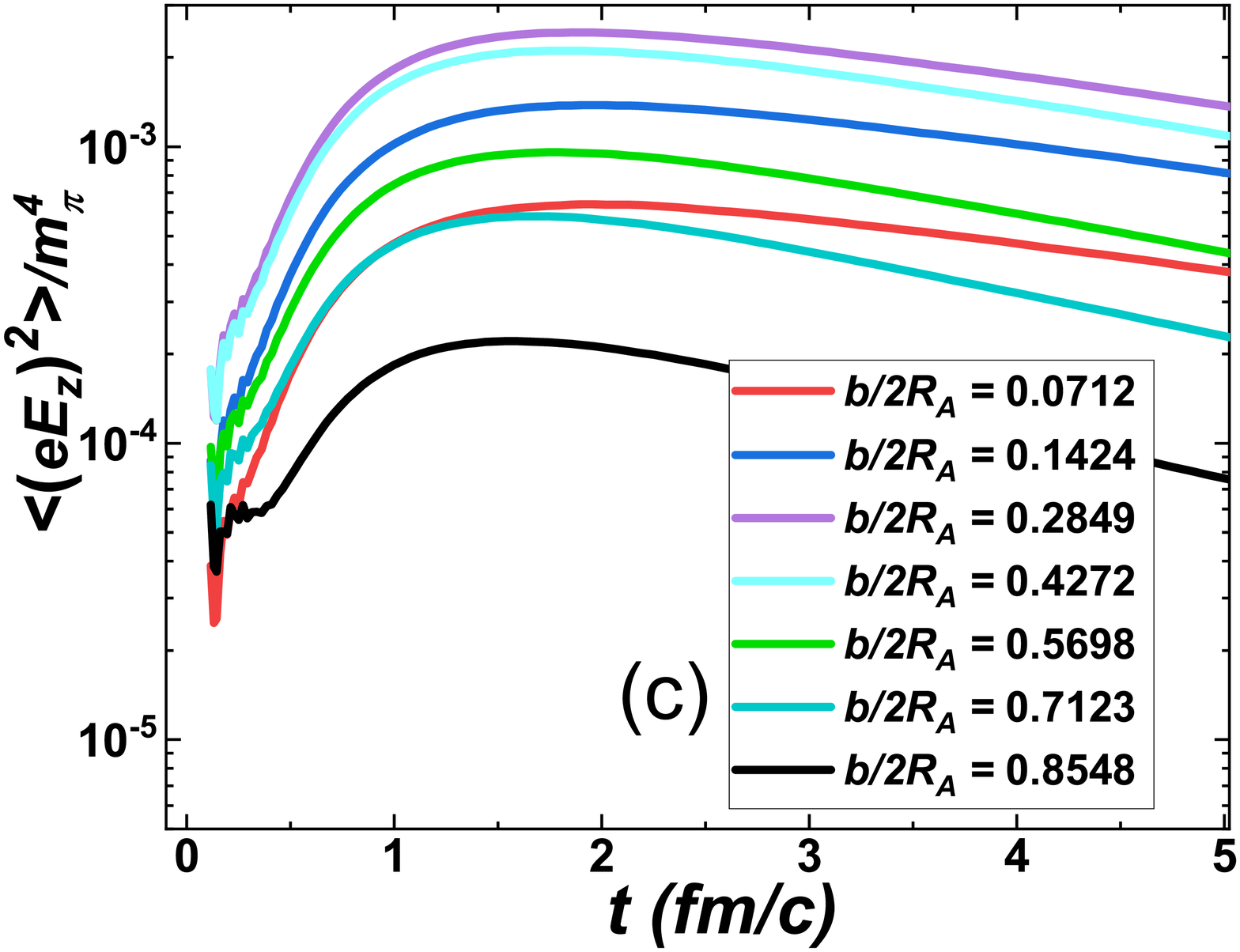}
\par\end{centering}
\caption{\label{fig:SquaredElectricfields_imp}Time evolution of $\left\langle (eE_{i})^{2}\right\rangle _{E}$
with $i=x,y,z$ for various impact parameters in Au+Au collisions
at $200$ GeV.}
\end{figure}

The results of the impact parameter dependence of $\left\langle (eB_{i})^{2}\right\rangle _{E}$
and $\left\langle (eE_{i})^{2}\right\rangle _{E}$ are given in Figs.
\ref{fig:SqauredFields_imp} and \ref{fig:SquaredElectricfields_imp}
respectively. The collision energy is set to 200 GeV and the impact
parameter is set to $b=1,2,4,6,7,10,12$ fm. We see in Figs. \ref{fig:SqauredFields_imp}(a) and (b) that $\left\langle (eB_{x})^{2}\right\rangle _{E}$
and $\left\langle (eB_{y})^{2}\right\rangle _{E}$ reach their maximum
values at about $t=0.08$ fm/c and fall fastly towards zero after
$t=1$ fm/c. We observe that $\left\langle (eB_{y})^{2}\right\rangle _{E}$
increases with the impact parameter, similar to $\left\langle eB_{y}\right\rangle _{E}$
. However, the peak values of$\left\langle (eB_{x})^{2}\right\rangle _{E}$
reach a maximum at an intermediate impact parameter. Such a non-monotonous
behaviour in the maximum values of $\left\langle (eB_{x})^{2}\right\rangle _{E}$
reflects charge fluctuations in the fireball. For small impact parameters,
fluctuations are relatively small comparing with large average charge
densities in the collision zone. For large impact parameters, fluctuations
are suppressed because of low energy densities in the collision zone.
The values of $\left\langle (eB_{z})^{2}\right\rangle _{E}$, as shown
in Fig. \ref{fig:SqauredFields_imp}(c), are about
three and four orders of magnitude smaller than $\left\langle (eB_{x})^{2}\right\rangle _{E}$
and $\left\langle (eB_{y})^{2}\right\rangle _{E}$ respectively, because
the $z$-component of the magnetic field is suppressed by the Lorentz
factor for particles moving in the $z$-direction. We also see the
peak values of $\left\langle (eB_{z})^{2}\right\rangle _{E}$ reach
a maximum at an intermediate impact parameter.

Similar impact parameter dependences also exist for squared electric
fields, $\left\langle (eE_{i})^{2}\right\rangle _{E}$, as shown in
Fig. \ref{fig:SquaredElectricfields_imp}. The maximum value of $\left\langle (eE_{x})^{2}\right\rangle _{E}$
appears at $b=2$ fm, while the maximum values of $\left\langle (eE_{y})^{2}\right\rangle _{E}$
and $\left\langle (eE_{z})^{2}\right\rangle _{E}$ appear at $b=4$
fm. The magnitudes of $\left\langle (eE_{x})^{2}\right\rangle _{E}$
and $\left\langle (eE_{y})^{2}\right\rangle _{E}$ are comparable,
which are about three orders of magnitude larger than $\left\langle (eE_{z})^{2}\right\rangle _{E}$.
For the impact parameter $b\geq7$ fm, there are two peaks in $\left\langle (eE_{x})^{2}\right\rangle _{E}$
as functions of time. This is because $E_{x}$ generated by the fireball
and spectators cancel in some space-time region. For small impact
parameters, $E_{x}$ generated by the fireball is significantly larger
than that by spectators, thus the second peak disappears.

\section{EM anomaly \label{sec:EM-anomaly}}

\begin{figure}[tb]
\begin{centering}
\includegraphics[width=14cm]{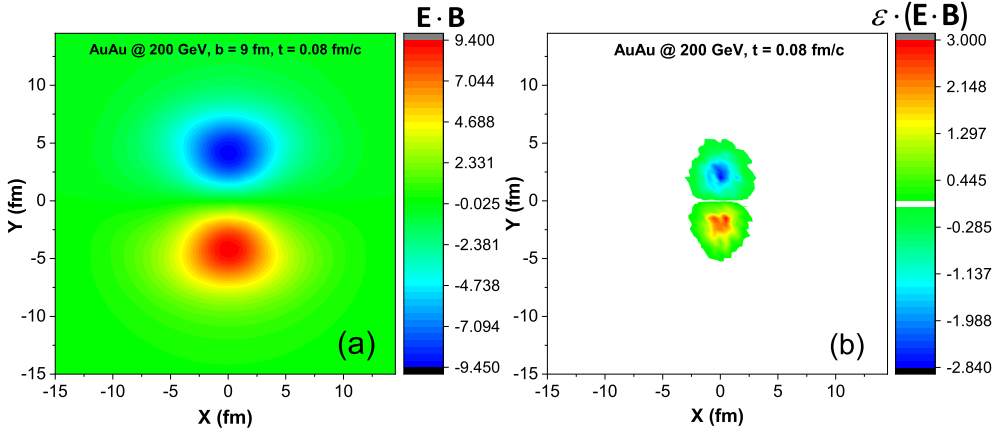}
\par\end{centering}
\caption{\label{fig:Spatial_E_B}The spatial distributions of $e^{2}\mathbf{E}\cdot\mathbf{B}$
[(a)] and the product of the energy density and $e^{2}\mathbf{E}\cdot\mathbf{B}$
[(b)].}
\end{figure}

\begin{figure}[h]
\begin{centering}
\includegraphics[width=8cm,height=6cm]{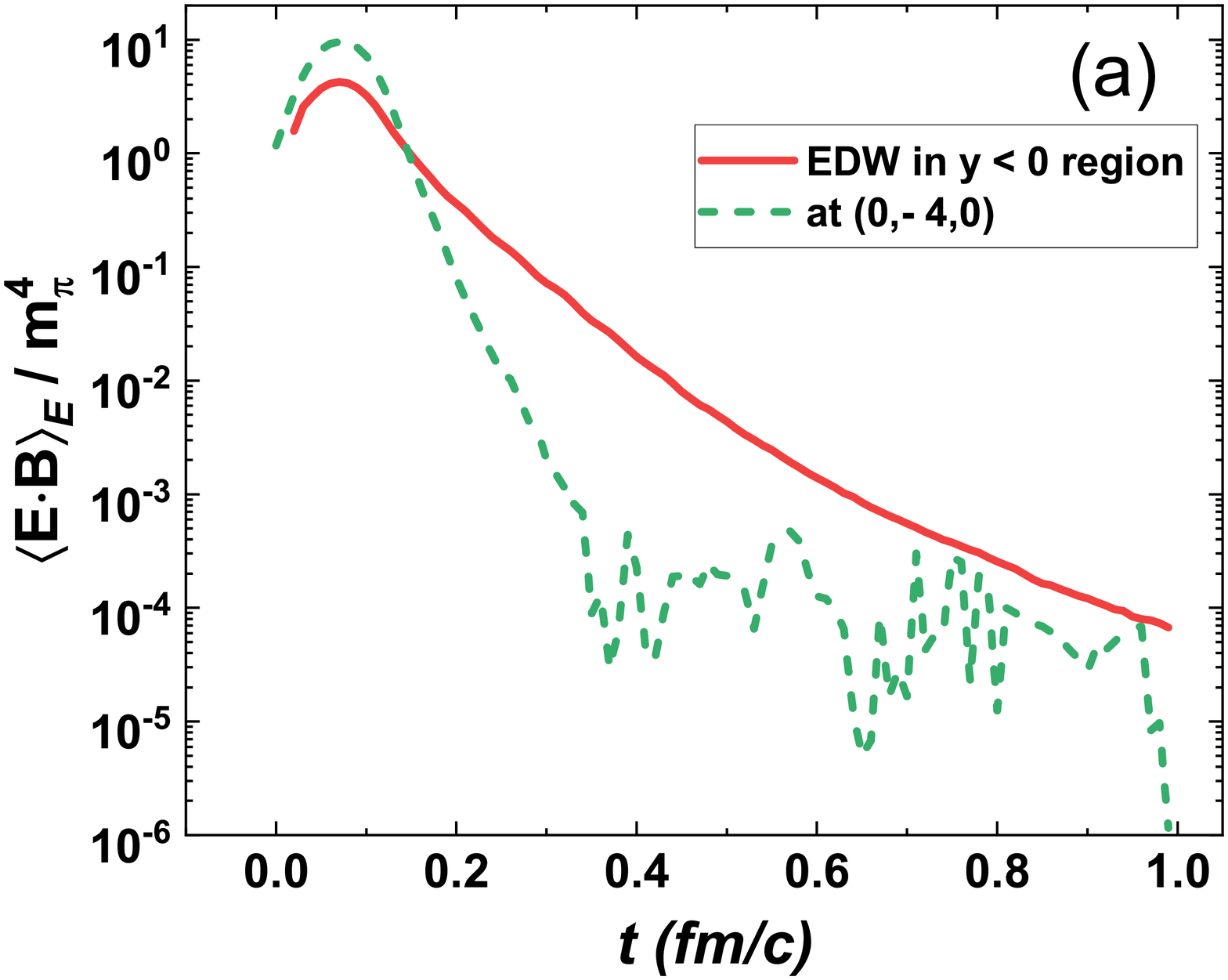}\includegraphics[width=8cm,height=6cm]{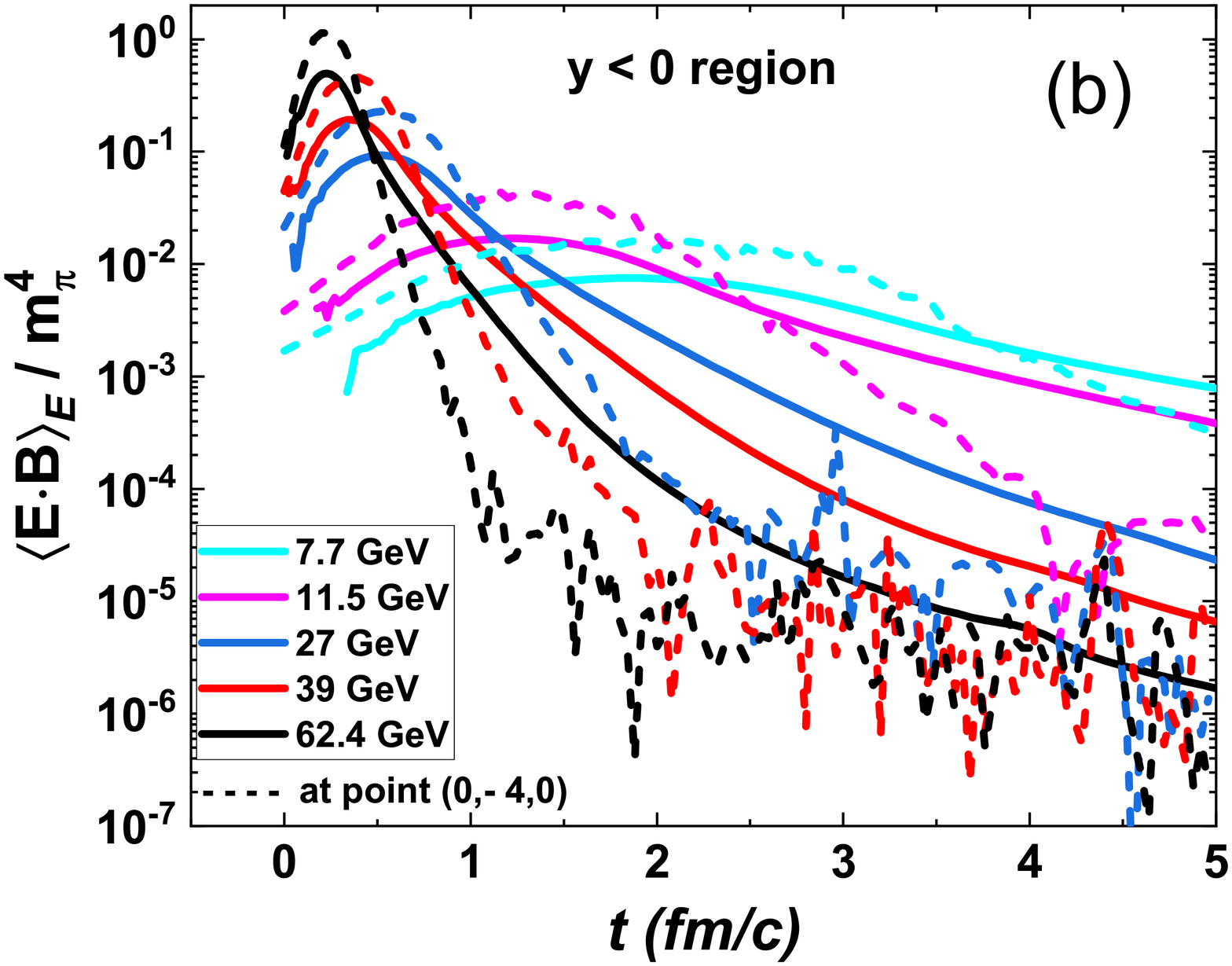}
\par\end{centering}
\caption{\label{fig:E_B_up_down}The time evolution of the space-average EM
anomaly weighted by the energy density $\left\langle e^{2}\mathbf{E}\cdot\mathbf{B}\right\rangle _{E}$
in the $-y$ region in Au+Au collision at 200 GeV [(a)] and
lower energies [(b)] and $b=9$ fm. The value of $e^{2}\mathbf{E}\cdot\mathbf{B}$
at the space point $(0,-4,0)$ fm as a function of time is also shown
for a comparison.}
\end{figure}

In this section we study the EM anomaly $e^{2}\mathbf{E}\cdot\mathbf{B}$
in heavy ion collisions. The spatial distribution of $e^{2}\mathbf{E}\cdot\mathbf{B}$
at $t=0.08$ fm/c for Au+Au collisions at 200 GeV and $b=9$ fm is
shown in Fig. \ref{fig:Spatial_E_B}. We choose $t=0.08$ fm/c because
the magnetic field reaches its maximum value at this time as shown
in Fig. \ref{fig:MagneticField-ImpactParameter}. The anomaly $e^{2}\mathbf{E}\cdot\mathbf{B}$
is symmetric for flipping the sign of $x$ and anti-symmetric for
flipping the sign of $y$, i.e. it is a dipolar distribution. Figure \ref{fig:Spatial_E_B} (b) shows the spatial distribution
of $e^{2}\mathbf{E}\cdot\mathbf{B}$ times the energy density, which
also has a dipolar structure. When directly calculating the space-average
of the EM anomaly weighted by the energy density, it is natural to
see $\left\langle e^{2}\mathbf{E}\cdot\mathbf{B}\right\rangle _{E}=0$,
but the averages in upper ($+y$) and lower ($-y$) half space are
all nonzero. In Fig. \ref{fig:E_B_up_down}, we show $\left\langle e^{2}\mathbf{E}\cdot\mathbf{B}\right\rangle _{E}$
as functions of time in the $-y$ region at 200 GeV [Fig. \ref{fig:E_B_up_down}(a)] and
lower energies 62.4, 39, 27, 11.5, 7.7 GeV [Fig. \ref{fig:E_B_up_down}(b)], and a comparison
has been made between $\left\langle e^{2}\mathbf{E}\cdot\mathbf{B}\right\rangle _{E}$
and $e^{2}\mathbf{E}\cdot\mathbf{B}$ at the space point $(0,-4,0)$
fm at each energy. Comparing with $e^{2}\mathbf{E}\cdot\mathbf{B}$
at the space point $(0,-4,0)$ fm, $\left\langle e^{2}\mathbf{E}\cdot\mathbf{B}\right\rangle _{E}$
have smaller peak values and decrease slower in time.

\begin{figure}[h]
\begin{centering}
\includegraphics[width=8cm]{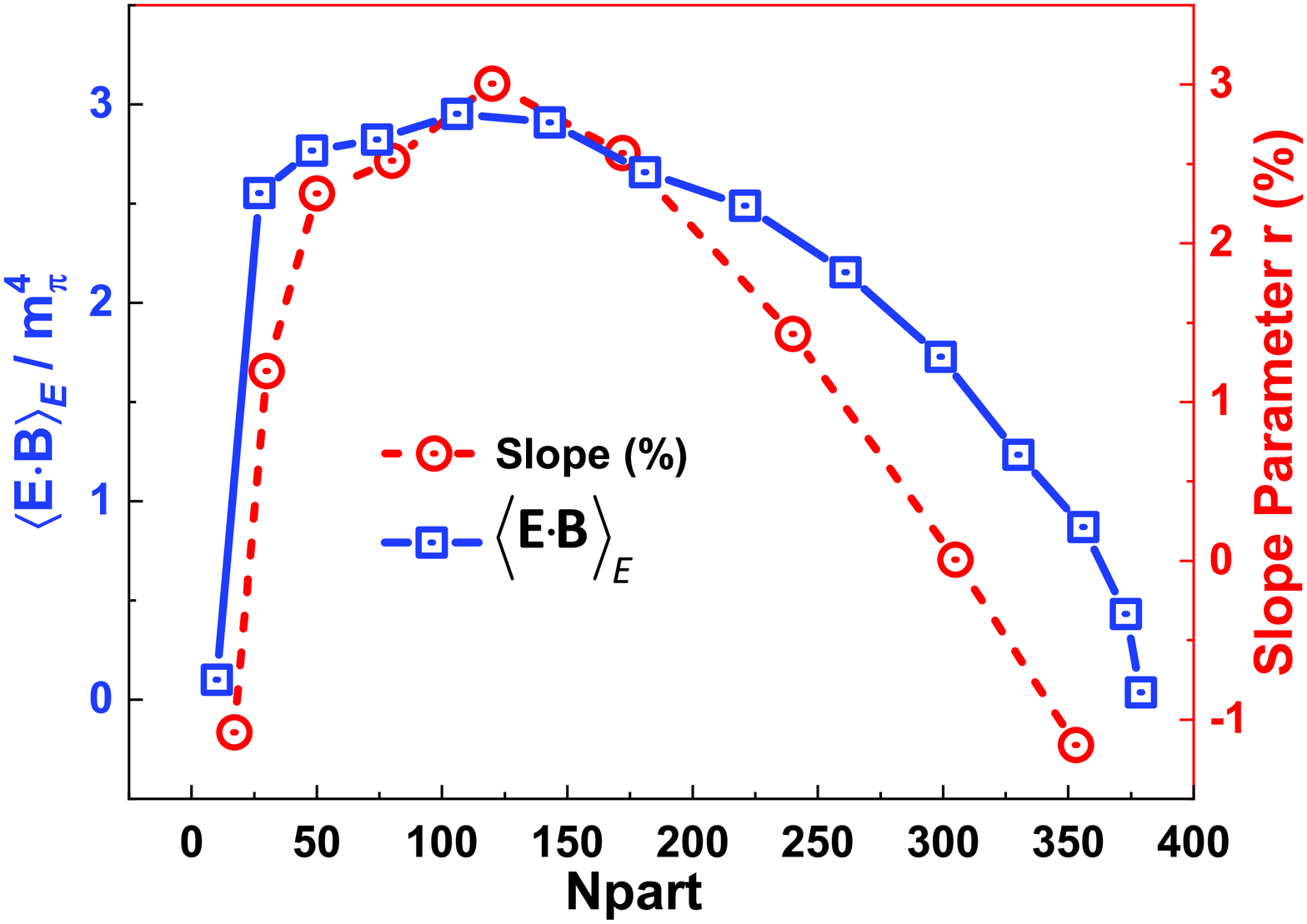}
\par\end{centering}
\caption{\label{fig:Centrality-dependence-E_B}The peak values of $\left\langle e^{2}\mathbf{E}\cdot\mathbf{B}\right\rangle _{E}$
and the slope parameter as functions of the number of participants
$N_{\text{part}}$ in Au+Au collisions at 200 GeV. }
\end{figure}

In Fig. \ref{fig:Centrality-dependence-E_B}, we give peak values
of $\left\langle e^{2}\mathbf{E}\cdot\mathbf{B}\right\rangle _{E}$
as a function of the number of participants at 200 GeV, compared with
the slope parameter for the difference in charge-dependent elliptic
flows for charged pions, which is measured by the STAR collaboration
\citep{Adamczyk:2015eqo}. We confirm that the $N_{\text{part}}$
dependence of $\left\langle e^{2}\mathbf{E}\cdot\mathbf{B}\right\rangle _{E}$
is consistent with that of the slope parameter. We note that $\left\langle e^{2}\mathbf{E}\cdot\mathbf{B}\right\rangle _{E}$
in the $+y$ and $-y$ region have an opposite sign, leading to opposite
chiral charges in the $\pm y$ regions and therefore a charge separation
with respect to the reaction plane because of the CME. Similar to
the CMW, this mechanism can also induce the charge-dependence $v_{2}$
observed in the STAR experiments \citep{Adamczyk:2015eqo,Zhao:2019ybo}.
Our results of $\left\langle e^{2}\mathbf{E}\cdot\mathbf{B}\right\rangle _{E}$
is about 50\% smaller than the values in Ref. \citep{Zhao:2019ybo}
because different methods are used when calculating zone-averages.

\section{Parameterization for space-average fields \label{sec:Analytical-formula-of}}

In previous sections we have presented results of space-average fields
for various collision energies and impact parameters. In this section,
we give parameterized formula for $\left\langle eB_{y}\right\rangle _{E}$,
$\left\langle (eB_{i})^{2}\right\rangle _{E}$, $\left\langle (eE_{i})^{2}\right\rangle _{E}$,
for $i=x,y$, and $\left\langle e^{2}{\bf E}\cdot{\bf B}\right\rangle _{E}$,
as functions of time. The other components $\left\langle eB_{x,z}\right\rangle _{E}$,
$\left\langle eE_{x,y,z}\right\rangle _{E}$, $\left\langle (eB_{z})^{2}\right\rangle _{E}$
and $\left\langle (eE_{z})^{2}\right\rangle _{E}$ are too small to
be parameterized. These analytical formulas are useful in studies
of field-related effects in heavy-ion collisions.

We notice from Fig. \ref{fig:MagneticField-ImpactParameter} and Fig.
\ref{fig:CollisionEnergyDependence} that $\left\langle eB_{y}\right\rangle _{E}$
as a function of time always has one peak at a specific time and the
peak value depends on both the impact parameter and the collision
energy, as shown in Figs. \ref{fig:MagneticField-ImpactParameter}(b)
and Fig. \ref{fig:CollisionEnergyDependence}(b). The average quantities
$\left\langle (eB_{x})^{2}\right\rangle _{E}$, $\left\langle (eB_{y})^{2}\right\rangle _{E}$,
$\left\langle (eE_{y})^{2}\right\rangle _{E}$, and $\left\langle e^{2}\mathbf{E}\cdot\mathbf{B}\right\rangle _{E}$
in Fig. \ref{fig:SqauredFields}, Fig. \ref{fig:SquaredElectricfields},
and Fig. \ref{fig:E_B_up_down} also have the one-peak structure similar
to $\left\langle eB_{y}\right\rangle _{E}$. For the time behaviors
of these quantities, we assume the following parameterization
\begin{equation}
\left\langle F\right\rangle _{E}(t)=M_{F}f\left[(\gamma t-t_{F})/(\text{fm/c})\right],\label{eq:fitting function}
\end{equation}
where $F$ represents $eB_{y}$, $(eB_{x})^{2}$, $(eB_{y})^{2}$,
$(eE_{y})^{2}$ or $e^{2}\mathbf{E}\cdot\mathbf{B}$, $M_{F}$ denotes
the maximum value of $\left\langle F\right\rangle _{E}$ with $t_{F}$
being its corresponding time multiplied by the Lorentz factor $\gamma=\sqrt{s_{NN}}/(2m_{p})$
with the proton mass $m_{p}$, and $f(x)$ is a function of dimensionless
variable $x$ and has the maximum value $1$ at $x=0$. We can further
parameterize $M_{F}$ and $t_{F}$ in second polynomials of $\gamma$
and the dimensionless impact parameter $\overline{b}=b/(2R_{A})$
with $R_{A}$ being the nuclear radius and $R_{A}=7.02$ fm for gold
nuclei,
\begin{eqnarray}
M_{F} & = & \alpha_{M}(1+c_{M}^{(1)}\gamma+c_{M}^{(2)}\gamma^{2})(1+c_{M}^{(3)}\overline{b}+c_{M}^{(4)}\overline{b}^{2}),\label{eq:M_F}\\
t_{F} & = & \alpha_{t}(1+c_{t}^{(1)}\gamma+c_{t}^{(2)}\gamma^{2})(1+c_{t}^{(3)}\overline{b}+c_{t}^{(4)}\overline{b}^{2}),\label{eq:t_F}
\end{eqnarray}
where the parameters $\alpha_{M}$, $c_{M}^{(i)}$, $\alpha_{t}$,
and $c_{t}^{(i)}$, $i=1,2,3,4$ are determined by fitting the peak
values of $\left\langle F\right\rangle _{E}(t)$. They vary for different
quantities of $F$, as shown in Table \ref{tab:Parameters-MF} and
Table \ref{tab:Parameters-tF}. Note that $\left\langle F\right\rangle _{E}$
reaches its maximum value at $t\simeq t_{F}/\gamma$ instead of $t=0$
at high energies, which is attributed to the finite size of the colliding
nuclei. We see in Table \ref{tab:Parameters-MF} that the peak value
of $\left\langle eB_{y}\right\rangle _{E}$ is proportional to $\gamma b$
at the leading order, similar to the behavior found in Ref. \citep{Deng:2012pc}
about the magnetic field at a specific space-time point. The deviation
from the linear behavior is described by second power terms of $\gamma$
and $b$. However, the peak values of squared fields $\left\langle eB_{x,y}^{2}\right\rangle _{E}$
and $\left\langle eE_{y}^{2}\right\rangle _{E}$ do not linearly depend
on $b$ at the leading order, this is because the average squared
fields are mainly dominated by fluctuations. We also see that the
peak value of $\left\langle e^{2}\mathbf{E}\cdot\mathbf{B}\right\rangle _{E}$
is linearly proportional to $b$ at the leading order, same as $\left\langle eB_{y}\right\rangle _{E}$.

\begin{table}[tbh]
\begin{tabular}{|c|c|c|c|c|c|}
\hline
$F$  & $\alpha_{M}$  & $c_{M}^{(1)}$  & $c_{M}^{(2)}$  & $c_{M}^{(3)}$  & $c_{M}^{(4)}$\tabularnewline
\hline
\hline
$eB_{y}$  & $6.266\times10^{-4}\,Z\gamma\bar{b}m_{\pi}^{2}$  & $2.123\times10^{-3}$  & $-3.855\times10^{-5}$  & $0.3890$  & $-0.6430$\tabularnewline
\hline
$(eB_{x})^{2}$  & $1.053\times10^{-8}\,Z^{2}\gamma^{2}m_{\pi}^{4}$  & $7.624\times10^{-4}$  & $-3.965\times10^{-6}$  & $4.018$  & $-4.954$\tabularnewline
\hline
$(eB_{y})^{2}$  & $1.446\times10^{-8}\,Z^{2}\gamma^{2}m_{\pi}^{4}$  & $4.135\times10^{-3}$  & $-7.086\times10^{-5}$  & $12.97$  & $6.791$\tabularnewline
\hline
$(eE_{y})^{2}$  & $5.592\times10^{-8}\,Z^{2}\gamma^{2}m_{\pi}^{4}$  & $-4.260\times10^{-4}$  & $-1.807\times10^{-5}$  & $2.089$  & $-3.278$\tabularnewline
\hline
$e^{2}{\bf E}\cdot{\bf B}$  & $1.399\times10^{-7}\,Z^{2}\gamma^{2}\bar{b}m_{\pi}^{4}$  & $2.033\times10^{-3}$  & $-5.523\times10^{-5}$  & $0.4472$  & $-1.213$\tabularnewline
\hline
\end{tabular}

\caption{\label{tab:Parameters-MF}The parameters in $M_{F}$ for various quantities
of $F$. Here $Z$ is the proton number of the colliding nuclei with
$Z=79$ for Au+Au collisions.}
\end{table}

\begin{table}[tbh]
\begin{tabular}{|c|c|c|c|c|c|}
\hline
$F$  & $\alpha_{t}$  & $c_{t}^{(1)}$  & $c_{t}^{(2)}$  & $c_{t}^{(3)}$  & $c_{t}^{(4)}$\tabularnewline
\hline
\hline
$eB_{y}$  & $5.681$  & $-1.368\times10^{-3}$  & $3.254\times10^{-5}$  & $0.3569$  & $-0.1061$\tabularnewline
\hline
$(eB_{x})^{2}$  & $7.239$  & $-3.820\times10^{-3}$  & $2.461\times10^{-5}$  & $-0.1821$  & $0.4701$\tabularnewline
\hline
$(eB_{y})^{2}$  & $5.186$  & $-1.840\times10^{-3}$  & $3.726\times10^{-5}$  & $0.8977$  & $-0.5478$\tabularnewline
\hline
$(eE_{y})^{2}$  & $8.043$  & $-1.930\times10^{-3}$  & $2.368\times10^{-5}$  & $0.06777$  & $0.09616$\tabularnewline
\hline
$e^{2}{\bf E}\cdot{\bf B}$  & $6.571$  & $-1.173\times10^{-3}$  & $2.738\times10^{-5}$  & $0.4465$  & $-0.3172$\tabularnewline
\hline
\end{tabular}

\caption{\label{tab:Parameters-tF}The parameters in $t_{F}$ for various quantities
of $F$.}
\end{table}

\begin{figure}[tbh]
\includegraphics[width=8cm]{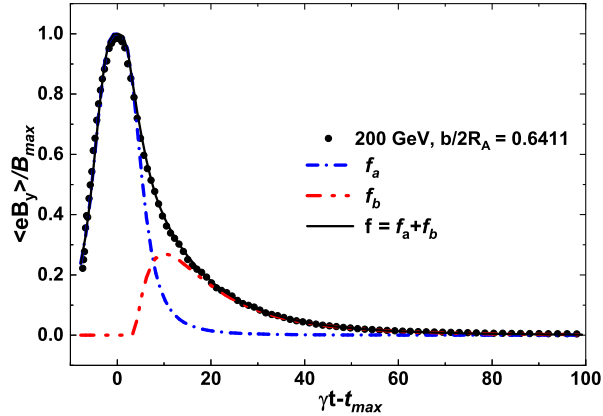}

\caption{\label{fig:Fitting}The function $f(x)$ for $\left\langle eB_{y}\right\rangle _{E}$
in Au+Au collisions at 200 GeV and $b=9$ in Eqs. (\ref{eq:function f},
\ref{eq:Fitting-function}).}
\end{figure}

The function $f(x)$ in Eq. (\ref{eq:fitting function}) can be further
written as a two-component form
\begin{equation}
f(x)=f_{a}(x)+f_{b}(x),\label{eq:function f}
\end{equation}
where $f_{a}(x)$ and $f_{b}(x)$ describe the early and later stage
of the evolution, respectively. We thus determine $f_{a}(x)$ by fitting
numerical results before the peak time and then determine $f_{b}(x)$
by fitting the difference between numerical results and $f_{a}(x)$.
The parameterization reads
\begin{eqnarray}
f_{a}(x) & = & \left[1+c_{a}^{(1)}(x^{2})^{c_{a}^{(2)}}\right]^{-1},\nonumber \\
f_{b}(x) & = & \theta(x-c_{b}^{(1)})\exp\left[c_{b}^{(2)}-c_{b}^{(3)}(x-c_{b}^{(1)})^{c_{b}^{(4)}}\right](x-c_{b}^{(1)})^{c_{P}^{(5)}},\label{eq:Fitting-function}
\end{eqnarray}
where $\theta(x)$ is the step function with $\theta(x>0)=1$ and
$\theta(x<0)=0$. The values of the parameters $c_{a}^{(i)}$ and
$c_{b}^{(j)}$ with $i=1,2$ and $j=1,2,3,4,5$ are given in Table
\ref{tab:Parameters-fx} which are determined by fitting the numerical
results of Au+Au collisions at $200$ GeV and $b=9$ fm. In Fig. \ref{fig:Fitting},
we plot $f_{a}$, $f_{b}$, and $f=f_{a}+f_{b}$ for $\left\langle eB_{y}\right\rangle _{E}$.
For comparison, we also show the numerical results for $\left\langle eB_{y}\right\rangle _{E}$
(black dots) from the UrQMD calculation. We see that $f_{a}$ dominates
at the early stage while $f_{b}$ dominates at the later stage as
expected.

\begin{table}[tbh]
\begin{tabular}{|c|c|c|c|c|c|c|c|}
\hline
$F$  & $c_{a}^{(1)}$  & $c_{a}^{(2)}$  & $c_{b}^{(1)}$  & $c_{b}^{(2)}$  & $c_{b}^{(3)}$  & $c_{b}^{(4)}$  & $c_{b}^{(5)}$\tabularnewline
\hline
\hline
$eB_{y}$  & $3.355\times10^{-3}$  & $1.609$  & $3.232$  & $11.20$  & $14.58$  & $0.2267$  & $5.217$\tabularnewline
\hline
$(eB_{x})^{2}$  & $7.744\times10^{-3}$  & $1.634$  & $2.167$  & $1.774$  & $4.897$  & $0.3830$  & $3.921$\tabularnewline
\hline
$(eB_{y})^{2}$  & $2.307\times10^{-3}$  & $1.918$  & $2.808$  & $6.659$  & $17.24$  & $-0.9004$  & $-2.867$\tabularnewline
\hline
$(eE_{y})^{2}$  & $0.02924$  & $1.294$  & $2.216$  & $-1.791$  & $2.163$  & $0.4969$  & $2.7374$\tabularnewline
\hline
$e^{2}{\bf E}\cdot{\bf B}$  & $7.744\times10^{-3}$  & $1.634$  & $4.046$  & $17.08$  & $20.29$  & $0.2147$  & $5.842$\tabularnewline
\hline
\end{tabular}

\caption{\label{tab:Parameters-fx}The parameters in $f(x)$ for various kinds
of $F$. }
\end{table}

The special quantity is $\left\langle (eE_{x})^{2}\right\rangle _{E}$,
which has two peaks as shown in Fig. \ref{fig:SquaredElectricfields},
different from $\left\langle eB_{y}\right\rangle _{E}$, $\left\langle (eB_{x,y})^{2}\right\rangle _{E}$,
$\left\langle (eE_{y})^{2}\right\rangle _{E}$, and $\left\langle e^{2}\mathbf{E}\cdot\mathbf{B}\right\rangle _{E}$.
We therefore parameterize $\left\langle (eE_{x})^{2}\right\rangle _{E}$
as
\begin{equation}
\left\langle (eE_{x})^{2}\right\rangle =M_{1}f_{a}\left[(\gamma t-t_{1})/(\text{fm/c})\right]+M_{2}f_{b}\left[(\gamma t-t_{2})/(\text{fm/c})\right]
\end{equation}
where $M_{1}$ and $t_{1}$ are for the first peak, while $M_{2}$
and $t_{2}$ are for the second peak. We assume the same parameterization,
Eqs. (\ref{eq:M_F}) and (\ref{eq:t_F}), for $M_{1,2}$ and $t_{1,2}$
as functions of $\gamma$ and $\overline{b}$. By fitting numerical
results, the parameters are obtained and given in Table \ref{tab:Parameters-M12}
and Table \ref{tab:Parameters-t12}. Meanwhile, $f_{a}$ and $f_{b}$
are also parameterized by Eq. (\ref{eq:Fitting-function}). Again,
the parameters in $f_{a}$ and $f_{b}$ are fixed by fitting $\left\langle (eE_{x})^{2}\right\rangle _{E}$
for Au+Au collisions at 200 GeV and $b=9$ fm, the results are given
in Table \ref{tab:Parameters-fafb}.

It is worthwhile to mention that the parameters in $f$ in Eq. (\ref{eq:Fitting-function})
can be determined by fitting numerical results at any collision energy
and any impact parameter with little difference although they are
determined in this paper by fitting numerical results at $200$ GeV
and $b=9$ fm. This means that $f$ is almost universal in a wide
range of collision energies from $7.7$ GeV to $200$ GeV and impact
parameters from $0$ to $2R_{A}$.

\begin{table}[tbh]
\begin{tabular}{|c|c|c|c|c|c|}
\hline
 & $\alpha_{M}$  & $c_{M}^{(1)}$  & $c_{M}^{(2)}$  & $c_{M}^{(3)}$  & $c_{M}^{(4)}$\tabularnewline
\hline
\hline
$M_{1}$  & $5.183\times10^{-8}\,Z^{2}\gamma^{2}m_{\pi}^{4}$  & $-3.054\times10^{-4}$  & $-9.345\times10^{-6}$  & $-0.8886$  & $0.05202$\tabularnewline
\hline
$M_{2}$  & $-5.908\times10^{-9}\,Z^{2}\gamma^{2}m_{\pi}^{4}$  & $1.153\times10^{-3}$  & $-2.619\times10^{-5}$  & $-14.17$  & $9.395$\tabularnewline
\hline
\end{tabular}

\caption{\label{tab:Parameters-M12}The parameters in $M_{1}$ and $M_{2}$
for $\left\langle (eE_{x})^{2}\right\rangle $.}
\end{table}

\begin{table}[tbh]
\begin{tabular}{|c|c|c|c|c|c|}
\hline
 & $\alpha_{t}$  & $c_{t}^{(1)}$  & $c_{t}^{(2)}$  & $c_{t}^{(3)}$  & $c_{t}^{(4)}$\tabularnewline
\hline
\hline
$t_{1}$  & $9.260$  & $-3.251\times10^{-3}$  & $2.653\times10^{-5}$  & $-0.4211$  & $-0.1054$\tabularnewline
\hline
$t_{2}$  & $6.935$  & $-2.135\times10^{-3}$  & $2.529\times10^{-5}$  & $3.238$  & $-2.170$\tabularnewline
\hline
\end{tabular}

\caption{\label{tab:Parameters-t12}The parameters in $t_{1}$ and $t_{2}$
for $\left\langle (eE_{x})^{2}\right\rangle $.}
\end{table}

\begin{table}[tbh]
\begin{tabular}{|c|c|c|c|c|c|c|c|}
\hline
 & $c_{a}^{(1)}$  & $c_{a}^{(2)}$  & $c_{b}^{(1)}$  & $c_{b}^{(2)}$  & $c_{b}^{(3)}$  & $c_{b}^{(4)}$  & $c_{b}^{(5)}$\tabularnewline
\hline
\hline
$\left\langle (eE_{x})^{2}\right\rangle _{E}$  & $0.02856$  & $1.356$  & $-15.81$  & $16.73$  & $235.6$  & $-1.547$  & $-4.926$\tabularnewline
\hline
\end{tabular}

\caption{\label{tab:Parameters-fafb}The parameters in $f_{a}$ and $f_{b}$
for $\left\langle (eE_{x})^{2}\right\rangle $.}
\end{table}

\section{Summary and conclusions\label{sec:Summary-and-conclusions}}

In this paper we use the UrQMD model to simulate the electromagnetic
fields in heavy ion collisions. In order to quantify the effects on
the hot and dense matter from electromagnetic fields, we propose the
space-average quantities (fields, squared fields, scalar product of
the electric and magnetic field, etc.) weighted by the energy or charge
density as functions of time to be barometers for field-related effects.
It is found that the average magnetic field increases with time and
reaches its maximum value soon after the collision, then it quickly
damps to zero. It is found that the peak value of the average magnetic
field is proportional to the collision energy and the impact parameter.
Comparing with the magnetic field at the geometric center of the collision,
the average quantities has a little smaller peak value shortly after
the collision but damps much slower or live much longer at the later
stage.

By fitting numerical results of electromagnetic fields with the UrQMD
model, we use analytical formula to parameterize the space-average
quantities, fields, squared fields, and electromagnetic anomaly (scalar
product of the electric and magnetic field), as functions of time.
The parameterization formulas are expressed in terms of the Lorentz
factor encoding the collision energy and the relative impact parameter
$\overline{b}=b/(2R_{A})$. We have checked that the parameterization
formulas are in good agreement with numerical results for collisions
at energies from 7.7 GeV to 200 GeV and impact parameters from 0 to
$12$ fm.

In the calculation of this paper, we do not introduce the electric
conductivity which is expected to slow down the damping of electromagnetic
fields and deserves a detailed study in the future.
\begin{acknowledgments}
The authors thank L. Oliva and X.-N. Wang for helpful discussions.
X.-L. S. is supported by the National Natural Science Foundation of
China (NSFC) under grants 11935007, 11221504, 11861131009, 11890714
(a sub-grant of 11890710) and 12047528. I.S. and Q.W. are supported in part by the National Natural
Science Foundation of China (NSFC) under Grants 11890713 (a sub-grant
of 11890710) and 11947301, and by the Strategic Priority Research
Program of Chinese Academy of Sciences under Grant XDB34030102.
\end{acknowledgments}

 \bibliographystyle{apsrev}
\bibliography{Eweighted}

\end{document}